\documentclass[aps,prb,showpacs,raggedbottom,nobalancelastpage,amssymb,twocolumn]{revtex4}
\usepackage{amsmath}
\usepackage{amsfonts}
\usepackage{graphicx}
\usepackage{dsfont}
\usepackage{amssymb}
\usepackage{bm}
\usepackage{color}
\usepackage{soul}
\usepackage{natbib}
\usepackage{array}
\usepackage{subfigure} 

\usepackage{natbib}
\usepackage{array}

\newcommand{\be}{\begin{equation}}
\newcommand{\ee}{\end{equation}}

\newcommand{\average}[1]{\left\langle#1\right\rangle}

\newcommand{\syse}{\end{array}\right.}

\newcommand{\ag}[1]{}

\usepackage{epsfig}
\usepackage{amssymb,amsmath,bbm,amsfonts,amsthm,subfigure,dsfont,eqnarray}
\usepackage{makecell}
\usepackage{setspace}

\newcommand{\abs}[1]{\left\vert#1\right\vert}

\begin{document}

\title{Universal scaling  of a classical impurity in the quantum Ising chain}

\author{Tony J. G. Apollaro$^{1,6}$}
\author{Gianluca Francica$^{2,3}$}
\author{Domenico Giuliano$^{2,3}$}
\author{Giovanni Falcone$^{2,3}$}
\author{G. Massimo Palma$^{4,5}$}
\author{Francesco Plastina$^{2,3}$}

\affiliation{$^1$ Quantum Technology Lab, Dipartimento di Fisica, Universit$\grave{a}$ degli Studi di Milano, 20133 Milano, Italy}
\affiliation{$^2$
Dipartimento di Fisica, Universit\`a della Calabria Arcavacata di
Rende I-87036, Cosenza, Italy} \affiliation{$^3$ I.N.F.N., Gruppo
collegato di Cosenza, Arcavacata di Rende I-87036, Cosenza, Italy}
\affiliation{$^4$ Dipartimento di Fisica e Chimica, Universit$\grave{a}$  degli Studi di Palermo, via Archirafi 36, I-90123 Palermo, Italy}
\affiliation{$^5$NEST, Istituto Nanoscienze-CNR }
\affiliation{$^6$Centre for Theoretical Atomic, Molecular and Optical Physics,
School of Mathematics and Physics, Queen's University Belfast, Belfast BT7 1NN, United Kingdom}

\begin{abstract}

We study finite size scaling for the magnetic observables of an
impurity residing at the endpoint of an open quantum Ising chain
with transverse magnetic field, realized by locally rescaling the
field by a factor $\mu \neq 1$. In the homogeneous chain limit at
$\mu = 1$, we find the expected finite size scaling for the
longitudinal impurity magnetization, with no specific scaling for
the transverse magnetization. At variance, in the classical
impurity limit, $\mu = 0$, we recover finite scaling   for the
longitudinal magnetization, while the transverse one basically
does not scale. We provide both analytic approximate expressions
for the magnetization and the susceptibility as well as numerical
evidences for the scaling behavior. At intermediate values of
$\mu$, finite size scaling is violated, and we provide a possible
explanation of this result in terms of the appearance of a second,
impurity related length scale. Finally, by going along the
standard quantum-to-classical mapping between statistical models,
we derive the classical counterpart of the quantum Ising chain
with an endpoint impurity as a classical Ising model on a square
lattice wrapped on a half-infinite cylinder, with the links along
the first circle modified as a function of $\mu$.

\end{abstract}

\pacs{75.10.Jm, 03.67.Bg}


\maketitle

\section{Introduction}\label{S.Intro}

Quantum phase transitions embody one of the most striking
collective behaviors of many body systems \cite{sachdev2011}. At
variance with thermal fluctuation induced phase transitions, a
quantum phase transition in a many body system is typically
triggered by quantum fluctuations and, therefore, it can take
place even at zero temperature, once a system parameter (say $h$)
is tuned across its critical value ($h_c$). In analogy with
thermal ones, the classification of quantum phase transitions
relies upon the Eherenfest-Landau scheme, by which one defines the
order of a transition as that of the lowest derivative of the
pertinent free energy functional, showing a discontinuity at the
critical point (see Ref.[\onlinecite{Jaeger1998}] for a review on
the subject). A milestone in the construction of a systematic
theory of the phase transitions is the concept of universality,
stating that all the physical systems sharing the same
dimensionality, symmetry of the order parameter and range of the
interaction, are expected to behave alike, close to a phase
transition and are said to belong to the same universality class
\cite{griffiths,kadanoff}. Universality is a consequence of the
divergence of the correlation length $\xi$  at the critical point
 \cite{kadanoff}. Indeed, near a second order quantum phase
transition, the growth of $\xi$ makes it the only relevant length
scale of the system, and makes the microscopic details of the
system irrelevant. Furthermore, it implies the scaling of physical
quantities as power laws of $ ( h_c - h )$, with critical
exponents that take the same values throughout the whole
universality class \cite{zinn_justin}. In addition, close to a
critical point, the algebraic divergence of $\xi$ implies scale
invariance of the system; that is, general physical quantities
behave as powers of control parameters times some scaling
functions of dimensionless ratios such as, for instance,
energy/(Boltzmann constant times) temperature, etc.
\cite{kadanoff,wilson}. An astonishing consequence of such a
prediction is that observables such as magnetization,
susceptibility, correlation length and time, specific heat,  as
well as  quantities which are not observables in the
quantum-mechanical sense, such as
entanglement~\cite{lorenzo_1,Osterloh2002, CampbellMDGAPBPNJP13},
Schmidt gap~\cite{PhysRevLett.109.237208}, irreversible
work~\cite{PhysRevB.93.201106}, all exhibit a scaling behavior
according to a set of critical indexes which define the
universality class the model belongs to.

Due to the recent progress in designing and fabricating quantum
devices with engineered properties, a remarkable interest has been
triggered in the physics of impurities in critical-, or
quasicritical-systems \cite{affleck08}. As an example, local
impurities have been proposed to improve the efficiency in quantum
state transfer protocols \cite{qst1,qst2}.

When impurities are realized in a critical system (the ``bulk''),
the lack of reference (energy- or length-) scales in the bulk
allows for the emergence of dynamically generated impurity-related
scales, such as the Kondo temperature, or the Kondo length, in the
case of magnetic impurities antiferromagnetically coupled to a
bulk of itinerant electrons \cite{kondo,hewson}, or of lattice
quantum spin systems \cite{furusaki98,sorensen}, or   the healing
length, in the case of tunnelling between interacting electronic
systems in one spatial dimension \cite{kane92}. Typically, the
impurity dynamics affects bulk quantities (such as the
conductance, or the spin susceptibility), and, in turn, it can be
probed by looking at the bulk response through suitably designed
devices. Recently, impurity induced dynamics has been
investigated, e.g., in Josephson junction networks
\cite{giuso0,giuso2,giuso3,giusox}, quantum spin chains
\cite{tsve1,tsve2,tsve3,gstt,gct, 2017arXiv170707838R}, and cold fermion gases
\cite{knap,schiro,sindona1,sindona2}.

While there is a remarkably large number of possible bulk effects
induced by the impurity dynamics, in this paper we take a
complementary point of view; namely, we rather look  at the
effects that a critical many-body bulk system has on the impurity.
In fact, several proposals have recently put forward to engineer
fully controllable quantum objects as ''quantum probes'' of many
body condensed matter systems (see, e.g.,
Ref.[\onlinecite{elliott,mitchison,tamascelli,streif,plastina_1}]).
In the specific context of a quantum impurity embedded within a
bulk system close to a quantum phase transition, either the
impurity generates a dynamic scale that rules the scaling of its
observables, or, if this does not happen, the impurity-related
observables scale with exponents that are directly linked to the
system ones. In the context of thermal phase transitions, this
effect has been demonstrated, e.g., for the surface magnetization
in inhomogeneous two-dimensional Ising lattices \cite{peschel_1}.
Due to the remarkable correspondence between $d+1$-dimensional
classical systems and quantum $d$-dimensional ones
\cite{fradkin_78,kogut_79}, one expects a similar behavior to
emerge for a quantum impurity embedded within one-dimensional
critical systems, namely that the scaling in the bulk implies some
sort of scaling in the impurity observables, as well
\cite{kim,0305-4470-18-1-006}.

To spell out the consequences of the bulk scaling on observable
quantities of a boundary impurity, here  we study the open,
one-dimensional, quantum Ising model in transverse field $h$
containing a side impurity. Since the seminal works dating back to
the '60~\cite{LIEB1961407,Barouch2}, the transverse field Ising
Model  has received a great attention in the literature;  its
critical indexes are since long well known~\cite{PhysRev.65.117,
PhysRev.85.808}, as well as experimentally verified (see
Ref.[\onlinecite{book:1390478}] and references therein). The
effects of  disorder and/or impurities have been considered as
well, unveiling quite a rich phenomenology that ranges from the
so-called Griffiths-McCoy singularities~\cite{PhysRevLett.23.17}
to the rounding of a quantum phase
transition~\cite{PhysRevLett.90.107202} (for an up-to-date review
of theoretical and experimental aspects see
Ref.[\onlinecite{1742-6596-529-1-012016}]).

In this paper, we consider a side impurity at the first site of
the Ising chain, realized with a local transverse magnetic field
equal to a fraction of the bulk one, $\mu h$ with $0 \leq \mu \leq
1$. At $\mu=1$, our model reduces back to the homogeneous one with
open boundary conditions. In contrast, for $\mu = 0$, we can
explore the physics of a classical impurity, see
Ref.[\onlinecite{nersesyan}]. As outlined there, a classical
impurity gives rise to a twofold degeneracy for the spectrum of
the whole system Hamiltonian. We show here that such an emerging
degeneracy results in the scaling behavior of the impurity
observables close to the bulk quantum phase transition.
Specifically, working at zero temperature, we will perform a
finite size scaling analysis of the physical properties of the
impurity and derive the corresponding scaling exponents. In
particular, we will investigate the finite size scaling behavior
of the impurity magnetization, by looking at both its longitudinal
and transverse components (directed along the coupling axis, and
along the applied magnetic field, respectively).

Close to the homogeneous chain limit ($\mu \sim 1$), we recover
the  finite size scaling of the longitudinal magnetization, which is consistent
with the behavior of the edge magnetization in the two-dimensional
classical model \cite{peschel_1}. At the same time, the transverse
magnetization shows no particular scaling properties. On the other
hand, when moving towards the classical impurity limit ($\mu
\rightarrow 0$), we find the emergence of finite size scaling in the transverse
magnetization, as well. While this appears to be an already
remarkable finding per se, it becomes particularly relevant when
interpreted along the results obtained in
Ref.[\onlinecite{apollaro}] where the energy spectrum is obtained
for the Hamiltonian of the inhomogenous transverse field Ising model.

The transverse field Ising model is known to exhibit a quantum
phase transition between an ordered phase and a paramagnetic one
\cite{sachdev2011}. Both phases are characterized by an excitation
spectrum with a finite energy gap, with spectra that appear quite
similar to each other, except for the appearance of a subgap mode
in the ordered phase, which eventually evolves towards an actual
zero-energy excitation as the size of the system increases. It is
exactly the appearance of the subgap mode that determines the
scaling behavior of the longitudinal impurity magnetization in the
ordered phase \cite{peschel_1} and, by converse, its absence which
determines the scaling to zero of the longitudinal magnetization
in the thermodynamic limit. The analogous contribution to the
transverse magnetization is, in general, overwhelmed by the
contributions from the modes with energy above the gap, which
yields no particular scaling behavior. In our inhomogeneous case,
when $\mu \to 0$, two important things happen: first, an
additional subgap mode emerges in the paramagnetic region and,
second, the contribution to the transverse magnetization from
above-the-gap modes shrinks to zero, thus providing the transverse
magnetization itself with an order parameter-like behavior
analogous to that of the longitudinal magnetization, but now in
the paramagnetic phase, rather than in the ordered one. As a
result, the behavior of the transverse magnetization close to the
quantum phase transition can be directly linked to the emergence
of such a subgap mode in the paramagnetic region, which is the
second, remarkable conclusion of our work.

The paper is organized as follows:

in Sec.~\ref{Ss.Isingdia}, for the sake of self-completeness, we
briefly recap the main results for the Ising model with open
boundary conditions and a single edge impurity reported in
Ref.[\onlinecite{apollaro}]; in Sec.~\ref{Ss.TM}, we report the
critical exponents of the impurity magnetic observables, which are
then used in Sec.~\ref{S.FSS} to obtain the predicted data
collapse. Finally in Sec.~\ref{S.XY} the universality hypothesis
is checked by verifying that the scaling exponents remain the same
also for the $XY$-model. Going along the correspondence between
$d$-dimensional quantum models and $d{+}1$-dimensional classical
statistical systems, in Sec.~\ref{cobc} we derive the classical,
two-dimensional analog of the transverse field Ising model with an
endpoint impurity. Finally, in Sec.~\ref{S.Concl} we provide our
main conclusions.

\section{Impurity model Hamiltonian}
\label{Ss.Isingdia}

We model the quantum impurity by rescaling the transverse magnetic field $h$ at
one endpoint of an $N$-site quantum Ising chain to $\mu h$, with $\mu$ being
a dimensionless parameter. Accordingly, our model Hamiltonian $H_\mu$ can be
regarded as a special case of the transverse field Ising model Hamiltonian in a non-uniform transverse
magnetic field $h_n$, that is

\begin{equation}
\hat{H}_\mu =
 -   J  \sum_{n=1}^{N -1} \hat{\sigma}^x_{n} \hat{\sigma}^x_{n+1} - J \sum_{ n =1}^N
h_n \hat{\sigma}^z_n  \;\;\;\;
, \label{E.HIsing}
\end{equation}
\noindent
with  $\hat{\sigma}^{\alpha}_n$
($\alpha=x,y,z$) being the Pauli matrices corresponding to (2 times) the components of a
 spin-1/2 quantum operator at site $n$ of the chain, $J$ being an over-all
energy scale (used as a reference scale  henceforth), and $ h_n =
h ( 1 - \delta_{n,1} ) + \mu h \delta_{ n ,  1} $. The model  in
Eq.(\ref{E.HIsing}) undergoes a quantum phase transition if the
magnetic field is set at the critical value $h_c=1$. In the
following, we will discuss the impurity physics for $h \sim h_c$.

\begin{figure}[ht]
       \begin{center}
         \includegraphics[width=\linewidth]{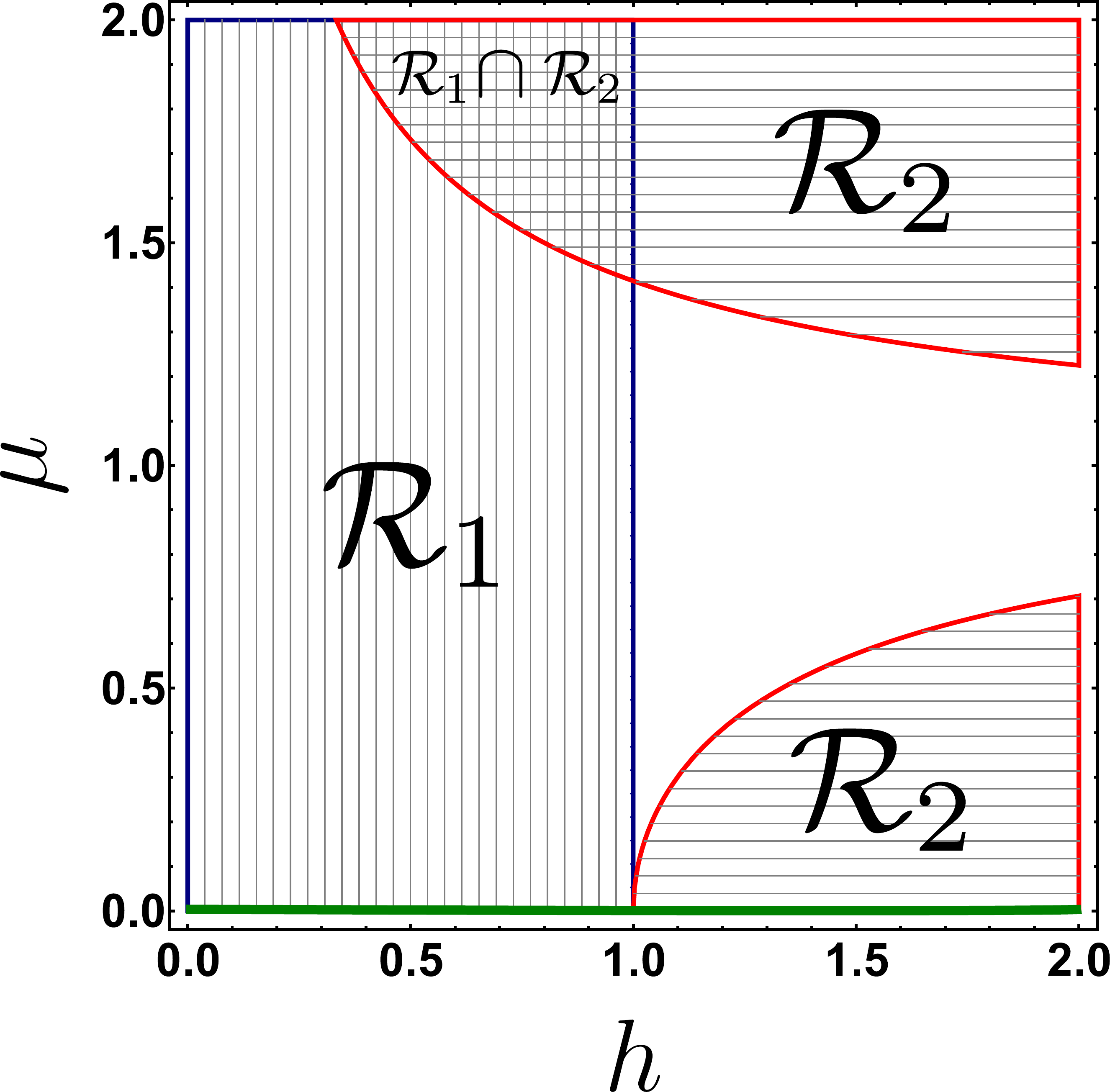}
     \caption{(Color online): Distribution of the discrete modes $\Lambda_{n}$ ($n{=}1,2$) in the $h,\mu$ plane.
        In the  region ${\cal R}_1$ (vertical lines), only the mode $\Lambda_1$ is present, while
        in the region ${\cal R}_2$ (horizontal lines), there is only the mode
        $\Lambda_2$,  either below or above the band in the two
        subregions with $\mu <1$, or  $\mu>1$, respectively.
        Both discrete modes are present in the intersection
        of ${\cal R}_1$ and ${\cal R}_2$.
        On the solid green line ($\mu=0$ with $h\geq 1$), mode-2 becomes an actual zero mode at
        any finite $N$ ($\Lambda_{2}=0$) and the impurity becomes classical \cite{nersesyan}.
         Finally, only quasicontinuous modes are present in
        the white region.}        \label{Figdiscr}
        \end{center}
\end{figure}

By the standard Jordan-Wigner representation of the spin-1/2
operators in terms of lattice spinless fermion operators $\{c_n ,
c_n^\dagger \}$, giving \cite{jordanwigner} $\hat{\sigma}^x_n =  [
c_n + c_n^\dagger ] \: e^{ i \pi \sum_{ r = 1}^{n-1} c_r^\dagger
c_r }$ and $\hat{\sigma}^z_n = 1 - 2 c_n^\dagger c_n$,
$\hat{H}_\mu $ is traded for an exactly solvable quadratic fermion
Hamiltonian. The derivation of the eigenvalues of $\hat{H}_\mu $
and of the corresponding eigenmodes is discussed in detail in
Ref.[\onlinecite{apollaro}]. Here, for the sake of the
presentation, we just review the main results. By introducing the
Bogoliubov-de Gennes quasiparticle operators $\{ \eta_q ,
\eta_q^\dagger \}$, related to the $\{ c_n , c_n^\dagger \}$ by
the Bogoliubov-Valatin transformations

\begin{eqnarray}
 \hat{\eta}_{q} &=&\sum_n \left\{  \frac{\psi_{q n}+\phi_{q n}}{2}\hat{c}_n
 +\frac{\psi_{q n}-\phi_{q n}}{2}\hat{c}^{\dagger}_n \right\} \nonumber
   \\
  \hat{\eta}_{q}^\dagger  &=&\sum_n \left\{  \frac{\psi_{q n}-\phi_{q n}}{2}\hat{c}_n
 +\frac{\psi_{q n}+ \phi_{q n}}{2}\hat{c}^{\dagger}_n \right\}
 \:\:\:\: .
 \label{bova.1}
 \end{eqnarray}
 \noindent
we can recast $\hat{H}_\mu$ in the form
 \begin{equation}
 \label{hdiag}
 \hat{H}_\mu= \sum_{\kappa}
\Lambda_{\kappa} \hat{\eta}_{\kappa}^{\dag} \hat{\eta}_{\kappa} + \chi_1
\Lambda_{1} \hat{\eta}_1^{\dag} \hat{\eta}_1 + \chi_2 \Lambda_{2}
\hat{\eta}_2^{\dag} \hat{\eta}_2 \, .
\end{equation}
\noindent In Eq.(\ref{hdiag}), $\kappa$ labels the quasicontinuous
modes, with energy larger than the single-quasiparticle bulk gap
$\Delta_m = | 1 - h |$: these are typically parametrized in terms
of the ''angles'' $\theta_\kappa$ which solve the secular
equations (A12) of Ref.[\onlinecite{apollaro}]. In addition,
$\hat{H}_\mu $ includes contributions from the two  {\it discrete}
modes $\eta_1 , \eta_2$, which, depending on the values of $h$ and
$\mu$, may appear at energies lying {\it within} $\Delta_m$. On
labelling with ${\cal R}_{n}$  the region in parameter space where
mode $n=1,2$ exists, we see that ${\cal R}_1$ is the ferromagnetic
region $h\leq 1$, while ${\cal R}_2=\{(h,\mu): (\forall h
\wedge|\mu|{>}\sqrt{1{+}1/h})\;{\vee}\;(h>1\wedge|\mu| {<}
\sqrt{1{-}1/h})\}$,  (${\cal R}_{1,2}$ are depicted in
Fig.~\ref{Figdiscr}; the details about their construction are
provided in Ref.[\onlinecite{apollaro}]). Furthermore, in Eq.
(\ref{hdiag}), we used $\chi_1 = \Theta (1-h)$ and $\chi_2$ to
denote the characteristic functions of ${\cal R}_1$ and ${\cal
R}_2$, respectively, so that the corresponding fermion mode $n=1$
($n=2$) is absent if $h, \mu$ are taken outside ${\cal R}_1$
(${\cal R}_2$). The wavefunctions $\psi_{q n} , \phi_{q n }$
entering Eq.(\ref{bova.1}), derived in
Ref.[\onlinecite{apollaro}], are reported here in Table
\ref{T.Sol}, where we list the energy eigenvalues of $\hat{H}_\mu$
and the corresponding eigenstates (in the $N \to \infty$-limit).

\begin{widetext}

\begin{table}[htp]
\caption{Expressions for the energy eigenvalues and the
$\{\psi,\phi\}$ matrix elements in the $(h,\mu)$-plane (see
Eq.(A12) of Ref.[\onlinecite{apollaro}] for the definition of the
parameter $\theta_\kappa$).}
\begin{center}
\begin{tabular}{|c|c|c|c|}
\hline
$(h,\mu)$ & $\Lambda$ & $\psi$ & $\phi$ \\
\hline
\thead{$\forall(h,\mu)$}
&
\thead{$\Lambda_\kappa{=} 2 \sqrt{ 1 {+} h^2 {-} 2 h \cos\theta_\kappa }$
}

&
\thead{$  \psi_n(\theta_\kappa){=} \sqrt{\frac{2}{N}} \frac{\sin(n\theta_\kappa) {+} (\mu^2-1)h
\sin((n{-}1)\theta_\kappa)}{\sqrt{1+(\mu^2-1)^2 h^2 {+} 2h (\mu^2 {-}1)\cos\theta_\kappa}}$ }
&
\thead{ $ \phi_n(\theta_\kappa){=}\frac{2h}{\Lambda_\kappa}  \psi_n(\theta_\kappa)
{-}\frac{2(\mu+1)h\delta_{n1}}{\Lambda_\kappa} \psi_1(\theta_\kappa){-}\frac{2(1{-}\delta_{n1})}{\Lambda_\kappa}
\psi_{n{-}1}(\theta_\kappa)$}
\\
\hline $\thead{{\cal R}_1}$ & $\thead{ \Lambda_{1} {=} \frac{2
|\mu|(1{-}h^2)h^{N}}{\sqrt{\abs{1{+}(\mu^2 {-}1) h^2 }}} }$ &
$\thead{ \psi^{(1)}_n {=} \sqrt{1{-}h^2}\left( h^{N{-}n} {-}
\frac{\mu^2 h^{N{+}n}}{1{+}(\mu^2-1)h^2} \right) }$ &

$\thead{\phi^{(1)}_n {=} \frac{\sqrt{1{-}h^2}\sqrt{\abs{1{+}(\mu^2-1)h^2}}}{|\mu| h(1{+}(\mu^2-1)h^2)}
\left( 1 {-} (\mu+1)\left( \delta_{n1}{+}(1{-}\delta_{n1})\left(1-\mu\right) \right) \right)h^n}$
\\
\hline \thead{${\cal R}_2$} & {\thead{$\Lambda_{2} = 2 |\mu|
\sqrt{\frac{1{+}(\mu^2-1)h^2}{(\mu^2-1)}}$}} &
\thead{$\psi^{(2)}_n{=} \frac{({-}1)^n
h^{-n}\sqrt{(\mu^2-1)^2h^{2}{-}1}}{{\left(\mu^2-1 \right)}^{n}}$}
& {\thead{$\phi^{(2)}_n {=}\frac{2h}{\Lambda^{(2)}}  \psi^{(2)}_n
{-}\frac{2(\mu+1)h \delta_{n1}}{\Lambda^{(2)}}  \psi^{(2)}_1  {-}
2\frac{1{-}\delta_{n1}}{\Lambda^{(2)}} \psi^{(2)}_{n{-}1}$}}\\
\hline
\end{tabular}
\label{T.Sol}
\end{center}
\end{table}
\end{widetext}

\section{Scaling analysis of the transverse and longitudinal magnetization in the thermodynamic limit}
\label{Ss.TM}

In this section we investigate the scaling  of the transverse and
longitudinal impurity magnetization in the vicinity of the bulk
quantum phase transition ($h \sim h_c = 1$) in the thermodynamic limit $N \to \infty$.
In addition,  we also  discuss the scaling  of the impurity
transverse susceptibility close to the critical point.

\subsection{Local transverse magnetization}
The transverse magnetization at site $n$ of the transverse field Ising model is defined as
the average value of the $z$ component of the spin operator at
that site, $\langle \hat{\sigma}_n^z \rangle$. Using the Jordan-Wigner
transformations  and inverting the
Bogoliubov-Valatin transformations in Eqs.(\ref{bova.1}), setting
$v_{qn}=\frac{\psi_{q n}-\phi_{q n}}{2}$, one obtains
\begin{equation}
 \langle \hat{\sigma}_n^z \rangle = 1 -2 \average{c_n^{\dagger}c_n}= 1- 2 \sum_q  v_{qn}^2
 \;\;\;\;.
 \label{tmag.1}
\end{equation}
\noindent When computing Eq.(\ref{tmag.1}) for  $n=1$ (transverse
impurity magnetization), we see that,  $\langle \hat{\sigma}_1^z
\rangle$ consists of various contributions, due to the
quasicontinuous modes $\lambda_\kappa$ and to the discrete modes
$\Lambda_{1,2}$. On singling out these three terms, one may write
down
\begin{equation}
 \langle \hat{\sigma}_1^z \rangle = \varphi_{\rm cm} + \varphi_1 + \varphi_2
 \:\:\:\: ,
 \label{tmag.2}
\end{equation}
\noindent
with
\begin{eqnarray}
 \varphi_{\rm cm}&{ =}& {-} \frac{4 \mu}{N} \: \sum_\kappa \: \left\{ \frac{\sin^2   \theta_\kappa   }{
 \Lambda_\kappa \: [ 1 {+} ( \mu^2 {-} 1 )^2 h^2 {+} 2 h ( \mu^2 {-} 1 ) \cos   \theta_\kappa  ] } \right\} \nonumber
   \\
 \varphi_1&{ =}& {-} 2 \left[ \frac{h^N{\rm sgn} ( \mu ) }{2 h } \right]\!
 \left[ \frac{( 1{ -} h^2 )^2}{\sqrt{1 {+} (\mu^2 {-} 1 ) h^2}} \right]{\rm sgn} [ 1 {+} ( \mu^2 {-} 1 ) h^2 ] \nonumber
\\
 \varphi_2& {=}& {-} 2 \left[ \frac{{\rm sgn} ( \mu ) }{h ( \mu^2 {-} 1 )^\frac{3}{2}} \right]
 \: \left[ \frac{ ( \mu^2 {-} 1 )^2 h^2 {-} 1 }{\sqrt{( \mu^2 {-} 1 ) h^2 {+} 1 }} \right]
 \:\:\:\: .
 \label{tmag.3}
\end{eqnarray}
\noindent
Eqs.~(\ref{tmag.2},\ref{tmag.3}) allow us to
separately discuss the large $N$ limit of the various
contributions to $\langle \hat{\sigma}_1^z \rangle$ close to the
quantum phase transition. First of all, we note that, in the homogenous
chain limit ($\mu \to 1$), $\varphi_2$ does not contribute
as  the mode $\Lambda_2$ is not present in the spectrum on the
horizontal line $\mu = 1$ in the parameter space, see
Fig.\ref{Figdiscr}. In this case, one simply obtains
\begin{equation}
 \langle \hat{\sigma}_1^z \rangle = -     h^{N-1} \: ( 1 - h^2 )^2    -
 \frac{4}{N} \: \sum_\kappa \: \left\{ \frac{\sin^2  \theta_\kappa  }{
 \Lambda_\kappa  } \right\}
 \;\;\;\; .
 \label{tmag.4}
\end{equation}
\noindent As $N \to \infty$, the contribution from the discrete
mode $\Lambda_1$ is exponentially suppressed, and one obtains
\begin{equation}
  \langle \hat{\sigma}_1^z \rangle \to
 - \frac{2}{\pi} \: \int_0^\pi \: d \theta \:  \left\{ \frac{\sin^2 \theta }{
 \sqrt{1 + h^2 - 2 h \cos \theta} } \right\}
 \:\:\:\: ,
 \label{tmag.5}
\end{equation}
\noindent
which is the standard result for the homogeneous transverse field Ising model.

\begin{figure}[ht]
        \begin{center}
        \includegraphics[width=.49\linewidth]{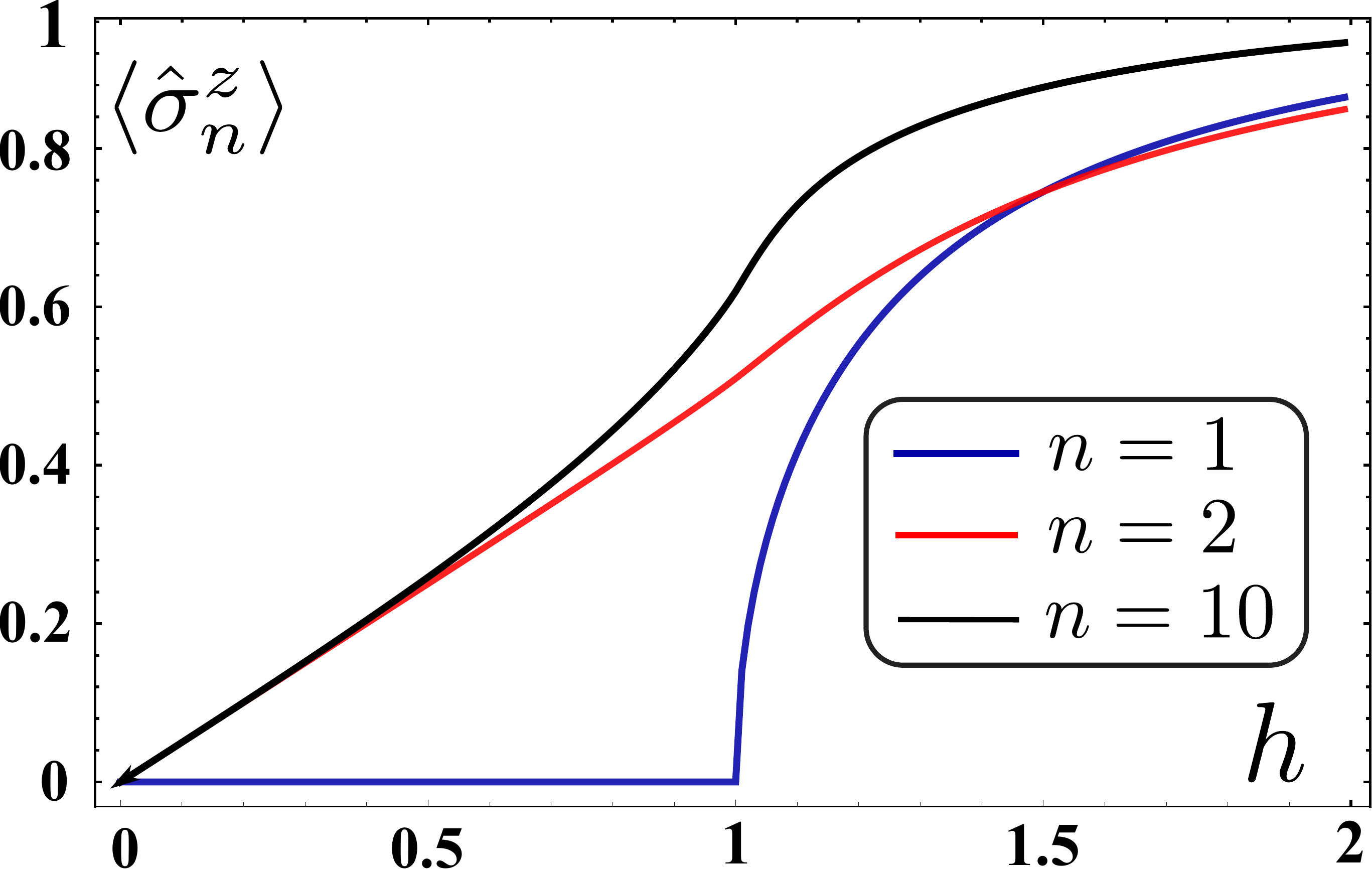}
        \includegraphics[width=.49\linewidth]{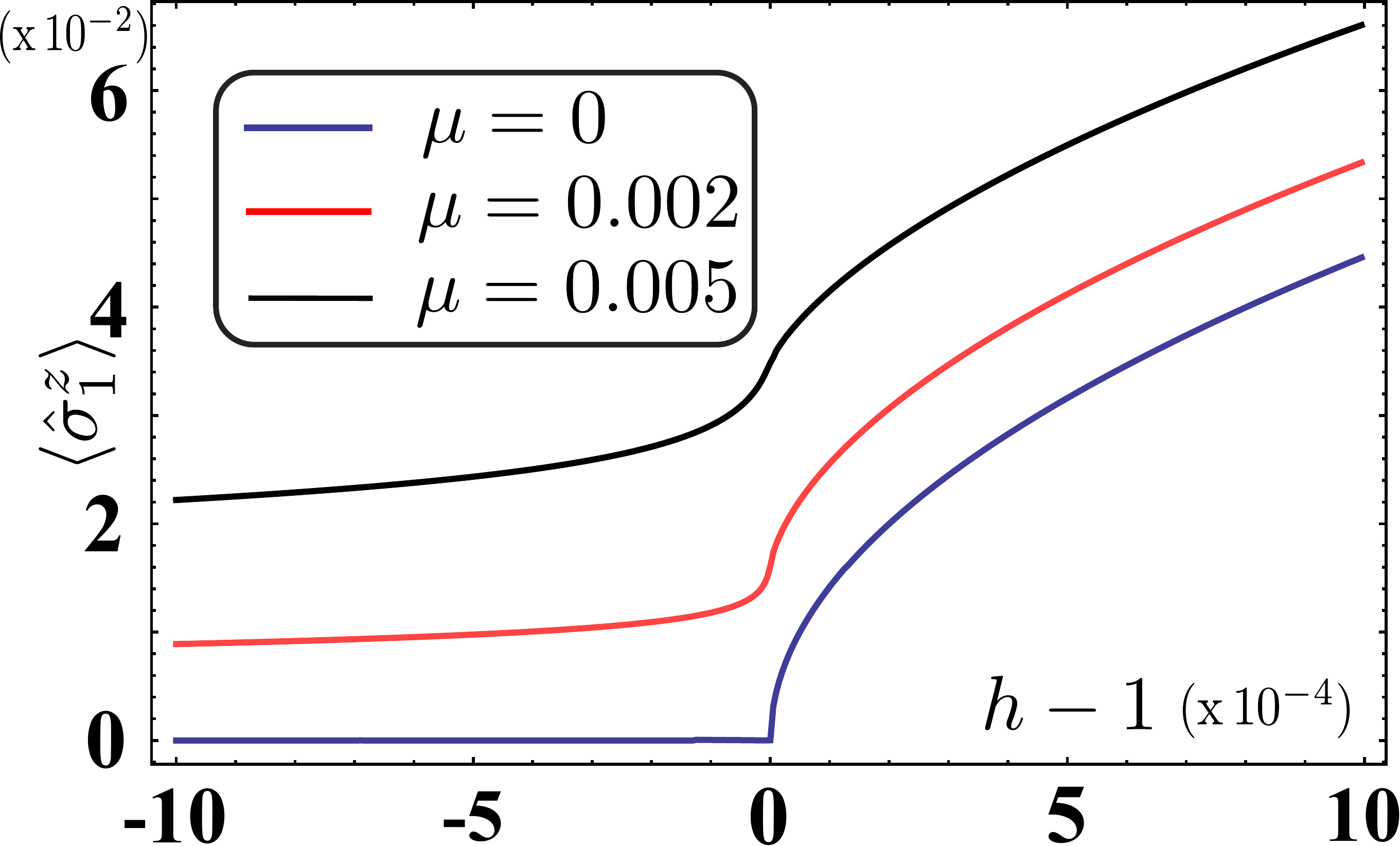}
\caption{(color online): \\  {\it Left Panel} - Transverse
magnetization $\langle {\hat{\sigma}}^z_n \rangle $ in the thermodynamic
limit ($N \to \infty$) as a function of the magnetic field $h$, as
$\mu\rightarrow 0$, for the impurity spin $n=1$ and, for
comparison, for spins $n=2$ and $n=10$ as well. Apparently, only
$\langle {\hat{\sigma}}^z_n \rangle $ behaves as a continuous
(dis-)order parameter as $h \to h_c$. \\
{\it Right Panel} - Transverse impurity magnetization $\langle
{\hat{\sigma}}^z_n \rangle$ (in the limit $N \to \infty$), near  $h_c$ and
for $\mu=0,0.002,0.005$. The step-like behavior holds only for
$\mu=0$.}\label{F.magzn}
        \end{center}
\end{figure}

At variance, when $\mu \to 0$, the discrete mode $\Lambda_2$
emerges with  vanishing energy  at any finite $N$. This
corresponds to the impurity at $n=1$ becoming classical
\cite{nersesyan}. Accordingly, the Hamiltonian spectrum becomes
twofold degenerate at any finite $N$ and one may always construct
a conserved operator commuting with $\hat{H}_{\mu = 0 }$, but
anticommuting with the Jordan-Wigner total fermion parity operator (see
Ref.[\onlinecite{nersesyan}] for a detailed discussion of this
point). Taking the $N \to \infty$ limit of Eq.(\ref{tmag.3}), one
eventually obtains
\begin{widetext}
\begin{equation}
\label{E.magn}
\langle \hat{\sigma}^z_1\rangle{=}
1 {-} 2 \left[\int_0^{\pi}d\theta\, v^2_1 (\theta){+}
\Theta\left(h{-}1\right)\Theta\left(x^+\right)
\Theta\left(y^+\right) \left(v_1^{(2)}\right)^2
{+}\Theta\left(1{-}h\right)\left( \left(\Theta\left(y^-\right){+}\Theta\left(x^-\right)\right) \left(v_1^{(2)}\right)^2{+}
\left(v_1^{(1)}\right)^2\right)\right]~,
\end{equation}
\end{widetext}
\noindent with  $x^{\pm}=\mu h + \sqrt{h (h\mp1)}$,
$y^{\pm}{=}-\mu h + \sqrt{h (h\mp1)}$, and $\Theta(x)$ being the
Heaviside step function. As $\mu \to 0$, Eq.(\ref{E.magn}) yields
\begin{equation}
 \label{E.mag1}
\displaystyle\lim_{\mu\to 0^{\pm}}\!\!\ \langle \hat{\sigma}^z_1 \rangle {=}
\mp \frac{\left(h+1\right)^{\frac{1}{2}}\left(h-1\right)^{\frac{1}{2}}}{ h}\Theta(h-1)~.
\end{equation}
\noindent
From Eq.(\ref{E.mag1}), we see that the transverse
magnetization has a power law behavior near $h_c$. Indeed, as $h
\to h_c$, one has $\langle \hat{\sigma}^z_1 \rangle \simeq 2
\left(h-1\right)^{\tilde{\beta}}$, with critical exponent
${\tilde{\beta}}=\frac{1}{2}$. Moreover, Eq.(\ref{E.mag1}) hints
towards a singular behavior of the $\mu \to 0$ impurity transverse
susceptibility $ \chi_z^{(1)} = \frac{\partial \langle {\hat{\sigma}}_1^z
\rangle }{\partial h}$ at $h \sim h_c$. Remarkably, such a
singular behavior of  $ \chi_z^{(1)} $ only  holds in the $\mu \to
0 $-limit, see Fig.\ref{F.magzn}. Note that the transverse
magnetization does not show any singularity in the homogeneous transverse field Ising model
($\mu = 1$), regardless on whether the boundary conditions are open or
periodic.
Now, from Eq.(\ref{E.mag1}), we see that, instead, as $\mu \to 0$,
$\langle \hat{\sigma}^z_1\rangle$ acquires a behavior similar to
what one would expect from an order parameter, except that it
takes a nonzero ground-state expectation value  in the
disordered paramagnetic phase ($h>h_c = 1$), rather than in the
ordered ferromagnetic one. This behavior is, in fact, a direct
consequence of the emergence of the discrete mode $\Lambda_2$ and
of the fact that it becomes a zero mode as $\mu \to 0$. The key
point is that, for all $\mu$, one has $ [ \hat{H}_\mu , \hat{P} ] =
0$, where $\hat{P}$ is the parity operator defined as $
\hat{P} \hat{\sigma}_n^{x , y} \hat{P} = - \hat{\sigma}_n^{x , y}$
and $\hat{P} \hat{\sigma}_n^z \hat{P} = \hat{\sigma}_n^z$. In
fermionic coordinates, $\hat{P}$ can be regarded as the operator
that changes the total fermion parity of a state,
\begin{equation}
 \hat{P} = \prod_{ n = 1}^N  \hat{\sigma}_n^z  = \prod_{ n = 1}^N \: e^{ i \pi c_n^\dagger c_n }
 \:\:\:\: ,
 \label{tmag.6}
\end{equation}
\noindent with the middle term of Eq.(\ref{tmag.6}) yielding the
realization of $\hat{P}$ in spin coordinates, while the last term
that for Jordan-Wigner-fermions. By construction, one has $\hat{P}
\eta_q = - \eta_q \hat{P}$, where $\eta_q$ is any one of the
fermion operators in Eq.(\ref{bova.1}). Therefore, the emergence
of a zero-energy $\eta_q$-mode is enough to ensure the twofold
degeneracy of the spectrum of $\hat{H}_\mu$. As it can be readily
inferred from Table \ref{T.Sol}, in a homogeneous chain ($\mu =
1$), this only happens for $h<1$ and for $N \to \infty$, where the
onset of the degeneracy implies a twofold degenerate ground state
for the system and, accordingly, the possibility for
$\hat{\sigma}_1^x$ to acquire a nonzero ground-state expectation
value, and to act as an order parameter for the ferromagnetic
phase.

In few words, in order to recover a nonzero ground-state
expectation value for a certain operator and use it as a non-zero
parameter to characterizes a given phase, one needs a degenerate
ground state to emerge in that phase. This is not the case for the
paramagnetic phase of the homogeneous transverse field Ising model
($\mu = 1$) which, indeed, does not show any nonzero order
parameter. On the contrary, when $\mu \to 0$, from
Eq.(\ref{E.HIsing}) one obtains $ [ \hat{\sigma}_1^x , \hat{H}_{
\mu = 0 } ] = 0$. Combining this last identity with the
observation that $ \hat{P} \hat{\sigma}_x^1 = - \hat{\sigma}_x^1
\hat{P}$, we eventually find two operators, both commuting with
$\hat{H}_{\mu = 0}$, but not commuting with each other. This is
enough to ensure the twofold spectral degeneracy of $\hat{H}_\mu$,
which is ultimately consistent with a nonzero ground-state
expectation value, $\langle \hat{\sigma}_1^z \rangle $, playing
the role of a characteristic parameter for such a phase.
Remarkably,  in this classical impurity limit, the degeneracy
emerges at any finite $N$, without the need of the thermodynamic
limit \cite{nersesyan}. In Jordan-Wigner fermion coordinates, the
twofold spectral degeneracy at $\mu = 0$ occurs when the discrete
mode $\eta_2$ becomes gapless: it corresponds to the zero-mode
operator in Eq.(14) of Ref.[\onlinecite{nersesyan}], adapted to
the case of a classical impurity at the endpoint of a
''semi-infinite'' chain. Another important point to stress here is
that, while in the ferromagnetic phase one gets $ \langle
\hat{\sigma}_n^x \rangle \neq 0$ for all $n$, both at the boundary
and in the bulk of the transverse field Ising model, see
Fig.\ref{F.magzn}, on the contrary, as $\mu \to 0$, only $\langle
\hat{\sigma}_1^z \rangle $ behaves like a (dis-)order parameter
for the paramagnetic phase, as $\langle \hat{\sigma}_n^z \rangle$
for $n > 1$ does not become zero when entering the ferromagnetic
phase $h<h_c$.

As for the impurity transverse susceptibility, for $N \to \infty$
and for $\mu \to 0$ one obtains $\chi_z^{(1)}=\frac{\Theta(h-1)}{2
h^2 \sqrt{h^2-1}}$. Thus, at the quantum phase transition, $\chi_z^{(1)}$ exhibits  an
algebraic divergence  ruled by the critical exponent
$\tilde{\alpha}=\frac{1}{2}$, at variance with what happens to the
transverse susceptibility in the bulk, which, at the quantum phase transition, diverges
only logarithmically, hence with critical exponent $\alpha=0$.

\subsection{Local longitudinal magnetization}
To discuss the behavior of the longitudinal magnetization of the
impurity, $\langle \hat{\sigma}_1^x \rangle $, in the
thermodynamic limit, one has to consider that, as discussed
before,  while $\hat{P}$ is always a symmetry of $\hat{H}_\mu$,
the twofold groundstate degeneracy is only recovered as $N \to
\infty$, if $\mu \ne 0$. Therefore, in performing the calculation
for finite-$N$ systems first, and then letting $N \to \infty$ (as
we do below, when discussing the impurity finite size scaling), one would
obtain $\langle \hat{\sigma}_1^x \rangle = 0$. To fix
this point, in the following  we adopt the method based on the
{\it{asymptotic factorizability}}. Specifically,  we compute
$\langle \sigma_1^x \rangle $ at any finite $N$ as
\begin{equation}
\label{E.Xfact}
\langle \hat{\sigma}^x_1 \rangle =\sqrt{\underset{r\rightarrow\infty}{\lim} \langle \hat{\sigma}^x_1 \hat{\sigma}^x_r \rangle }~,
\end{equation}
\noindent which is equivalent~\cite{RevModPhys.36.856} to
\begin{equation}
 \langle \hat{\sigma}_1^x \rangle \equiv | \langle E_0 | \hat{\sigma}^x_1 | E_1\rangle  | \equiv | h_1^{(0)}  |
 \:\:\:\: ,
 \label{sxx.1}
\end{equation}
with  $ |  E_0 \rangle$ and $| E_1\rangle $ being the ground and
first excited states of the transverse field Ising Hamiltonian at
given $N $ and $\mu$, respectively. Labelling the corresponding
eigenfunctions as $\psi_n^{(0)}$ and $\phi_n^{(0)}$, as in  Table
\ref{T.Sol}, one therefore obtains $h_1^{(0)} = \phi_1^{(0)}$. As
a result, if  $ |  E_0 \rangle$ and $| E_1\rangle $ become
degenerate as $N \rightarrow\infty$, giving rise to spontaneous
symmetry breaking, a nonzero value for $ \langle \hat{\sigma}_1^x
\rangle$ occurs. On the contrary, if a finite energy gap between
the two states persists even as $N \to \infty$, so that no
spontaneous symmetry breaking takes place, we have $ \langle
\hat{\sigma}_1^x \rangle = 0$. It is worth noticing that
Eq.(\ref{E.Xfact}) is actually valid only after the limiting
procedure (in fact, the exponent $\beta$ computed in
Refs.[\onlinecite{mccoy,peschel_1}] refers to the critical scaling
in a semi-infinite chain: no mention is made of finite size
scaling). Let's point out that Eq.(\ref{E.Xfact}) makes sense only
if the edge spins are exchangeable. While this is clearly true for
the homogenous transverse field Ising model, it does not apply to
our system at $\mu \neq 1$. Therefore, in order to correctly
evaluate $ \langle \hat{\sigma}_1^x \rangle$ with
Eq.(\ref{E.Xfact}), in performing the calculation at finite $N$,
we consider a  mirror-symmetric version of our system; that is, we
modify $h_n$ to $\tilde{h}_n$, given by
\begin{equation}
\tilde{h}_n = h ( 1 - \delta_{ n , 1 } ) ( 1 - \delta_{ n , N } )
+ \mu h \{ \delta_{ n , 1 } + \delta_{ n , N } \} \, ,
 \label{modfield}
\end{equation}
\noindent with a second, symmetric impurity at the end of the
chain. Apparently, this should yield the appropriate result for
our impurity system, provided one takes a large enough $N$. Taking
all the above caveats into account, and performing the calculation
using the wavefunctions in Table \ref{T.Sol}, one eventually
obtains that, as $N \to \infty$, the longitudinal impurity
magnetization is given by
\begin{equation}
\label{E.sigmax}
\langle \hat{\sigma}^x_1 \rangle
=\sqrt{\frac{1-h^2}{1+h^2\left(\mu^2-1\right)}}
\: \Theta\left(1-h\right) \; \; .
\end{equation}
\noindent Eq.(\ref{E.sigmax}) shows that, as expected, one gets
$\langle \hat{\sigma}^x_1 \rangle = 0$ for $h \geq 1$. At the same
time, for $h\rightarrow 1^-$,  one obtains to leading order
\begin{equation}
\label{E.sigmaxh1}
\langle \hat{\sigma}^x_1 \rangle \simeq \frac{\sqrt{2\left(1-h\right)}}{|\mu|} \:\:\:\:.
\end{equation}
\noindent From Eq.(\ref{E.sigmaxh1}) we infer that, on one hand, a
value of $\mu < 1$ gives rise to an effective renormalization of
the magnetic moment of the impurity spin \cite{nersesyan}. On the
other hand, and more importantly, the critical index
$\beta=\frac{1}{2}$ obtained  for the boundary magnetization in
the homogeneous open boundary transverse field Ising model \cite{PFEUTY197079,mccoy},  is
not affected for $\mu \neq 1$.

We now move to discussing the impurity finite size scaling in our inhomogenous transverse field Ising model.

\section{Finite-size impurity scaling}
\label{S.FSS}

The idea of finite size scaling~\cite{domb1983phase} is based on the observation that the behavior  of
a finite-size statistical system of typical  size $\sim a N$, with
$a$ being the lattice step (we set $a=1$ throughout this paper), is
determined by a scaled variable $x=\frac{N }{\xi}$, with $\xi $
being  the correlation length.
As scaling is a property of
large-size (i.e., thermodynamical) systems, one expects it to be
recovered only over a scale $x$ such that $x \ll 1$.
%

In this section we discuss  finite size scaling of the impurity
magnetization (transverse and longitudinal) as well as of the
transverse susceptibility eventually showing how, and for which
values of the system parameters, on varying the size $N$, the
corresponding data for the various impurity observables collapse
onto each other, with  scaling exponents consistent with those
derived in  Sec.~\ref{Ss.TM} in the thermodynamic limit.
Specifically, we first compute the various impurity-related
quantities at a finite system size $N$. Then, by fitting the
results obtained in this way with the standard finite size scaling
formulas, we extract the correlation length $\xi ( h )$ associated
to the bulk quantum phase transition as a function of $h$. We find
that
 $\xi(h) \sim \left| h- h_c\right|^{-\nu}$, with $\nu = 1$. Thus,  when either  $h\ll h_c$, or $h\gg h_c$,
(that is, far enough from the quantum phase transition), one finds
that  $\xi(h)\ll N$. Accordingly, in this regime the various
observable quantities  do not depend on $N$ and their values
correspond to what one finds in the thermodynamic limit. On the
other hand,  by approaching the quantum phase transition, $\xi(h)
\simeq N$, and, consequently,  the observables take a finite-size
dependence on $N$, which is the key point of our scaling
analysis~\cite{0305-4470-18-1-006,kim}.

\begin{figure}[ht]
        \begin{center}
  \includegraphics[width=\linewidth]{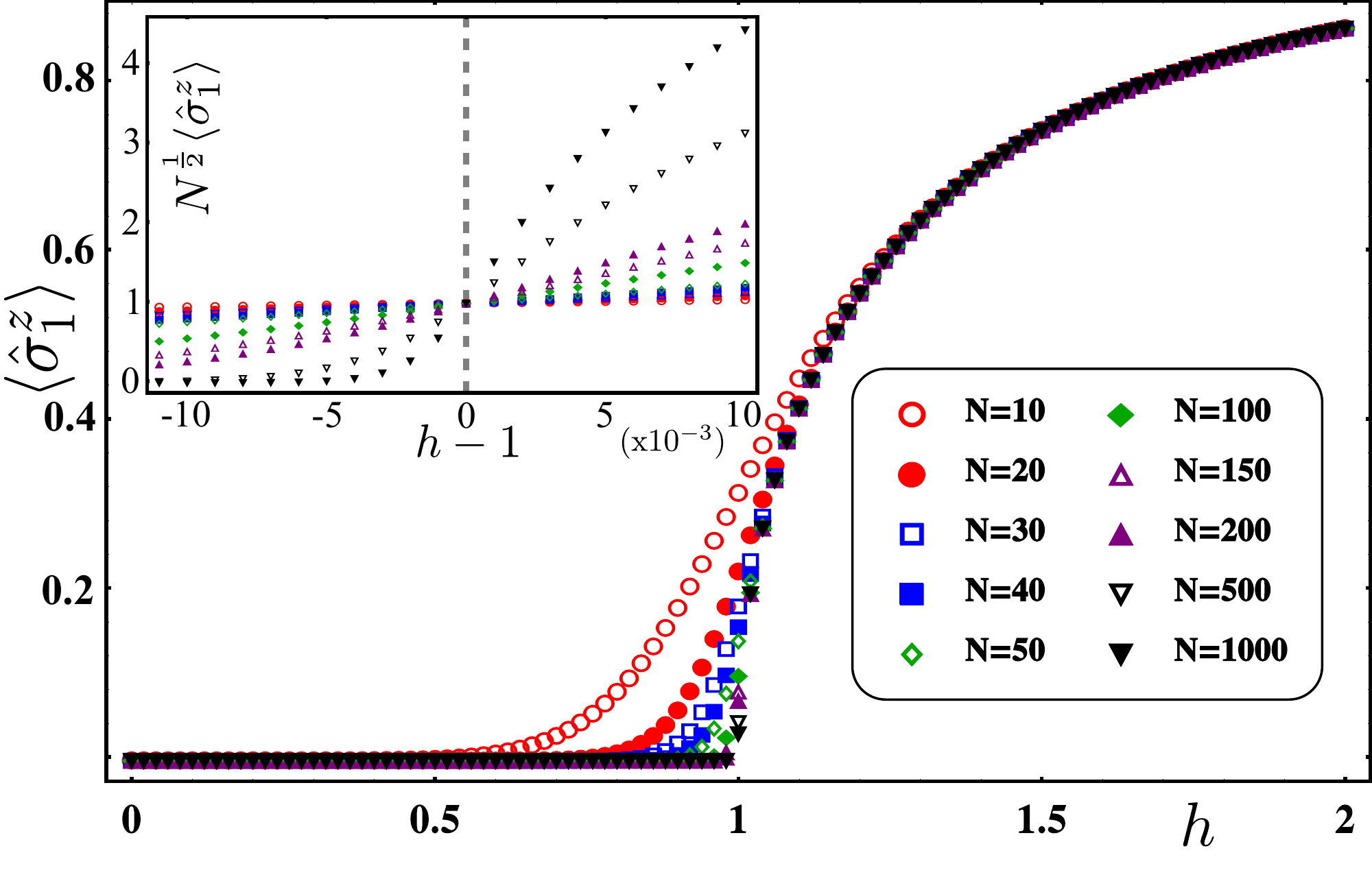}
        \includegraphics[width=\linewidth]{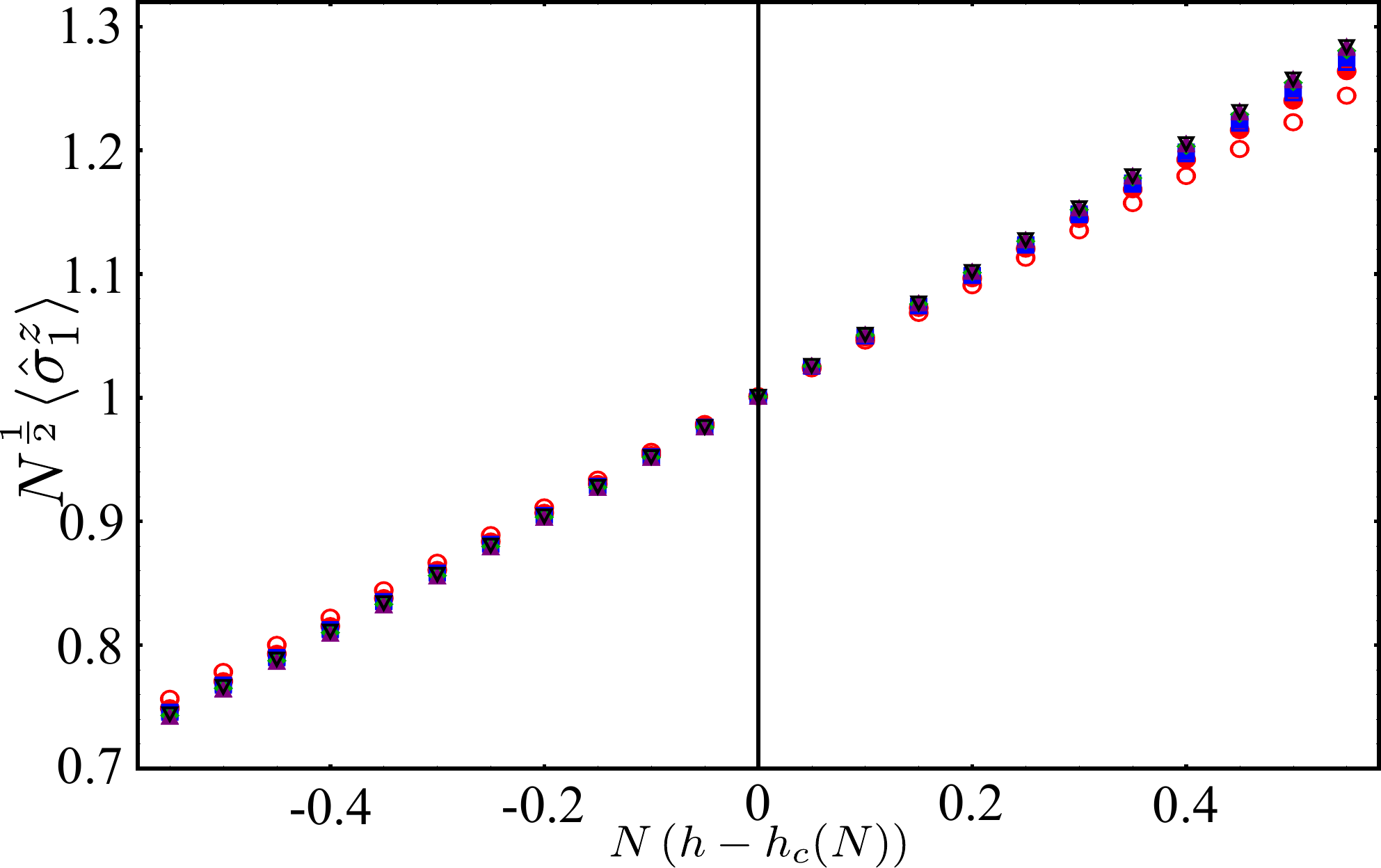}
\caption{(color online): {\it Upper plot} -  Transverse impurity
magnetization, $\langle \hat{\sigma}_1^z \rangle$, as a function
of the bulk magnetic field $h$ for different system sizes $N$.
{\it inset} - finite size scaling Ansatz
$N^{\frac{\tilde{\beta}}{\nu}}\average{\sigma_0^z}$ versus $h$,
with $\frac{\beta}{\nu}{=}\frac{1}{2}$. The intersection shows the
location of the critical point at $h_c{=}1$. {\it Lower plot} - Data collapse obtained with the scaling Ansatz of
Eq.~\ref{E.ScalingAnsatz}.  }\label{F.magz}
        \end{center}
\end{figure}

We now start to perform the finite size scaling analysis for the
transverse impurity magnetization close to the quantum phase
transition.  According to the derivation of Sec.~\ref{Ss.TM}, we
expect $\langle \hat{\sigma}_1^z \rangle$ to exhibit scaling only
for $ | \mu | \ll 1$. In this regime, the finite size scaling
ansatz for $\langle \hat{\sigma}_1^z \rangle$ gives
\begin{equation}
\label{E.ScalingAnsatz}
\langle \hat{\sigma}_1^z \rangle \sim N^{ - \frac{\tilde{\beta}}{\nu}}f\left(N^{\frac{1}{\nu}}\left| h-h'_c\right| \right) \:,
\end{equation}
\noindent with  the critical exponents $\tilde{\beta}$ and $\nu$,
describing  the singular behavior of the transverse impurity
magnetization and of the correlation length, respectively, and
with $f(x)$ being a suitable scaling function. According to
Eq.(\ref{E.ScalingAnsatz}), we determine the ratio
$\frac{\tilde{\beta}}{\nu}$  by plotting the rescaled
magnetization $N^{\frac{\tilde{\beta}}{\nu} } \langle {\hat{\sigma}}_1^z
\rangle$ versus $h$. In order to do so, recalling that, for $\mu
\to 0$, only the $\Lambda_2$-mode contributes to $\langle
\hat{\sigma}_1^z \rangle$ (see Eqs.(\ref{tmag.2},\ref{tmag.3})),
we have to employ the finite-$N$ version of the functions
$\psi_n^{(2)}$ and $\phi_n^{(2)}$ of Table \ref{T.Sol}, which, for
$\mu = 0$,  are
\begin{eqnarray}
 \psi_n^{(2)} = \sqrt{\frac{h^2-1}{1-h^{-2N}}} h^{-n}~~,~~\phi_n^{(2)} = \delta_{n 1} \;\;\;\; .\label{fn.1}
 \end{eqnarray}
\noindent This implies
\begin{equation}
\langle \hat{\sigma}^z_1 \rangle  =\frac{1}{2h}\sqrt{\frac{h^2-1}{1-h^{-2N}}}~,
\label{E.Sz}
\end{equation}
\noindent To first order in $\left(h-h_c\right)$, Eq.(\ref{E.Sz})
gives
\begin{equation}
\langle \hat{\sigma}^z_1 \rangle \approx  N^{- \frac{1}{2}}\left(\frac{1}{2}+\frac{N-1}{4}(h-1)\right)
\;\;\;\; .
\label{fn.2}
\end{equation}
\noindent From Eq.(\ref{E.Sz}) for $n=1$, we construct the plot
reported in the upper left inset of Fig.\ref{F.magz}, where, in
particular, we show that the curves representing the rescaled
magnetization versus $h$ intersect with each other at
$h{=}h_c{=}1$ once one sets
$\frac{\tilde{\beta}}{\nu}{=}\frac{1}{2}$. Having determined
$\frac{\tilde{\beta}}{\nu}$, we can proceed to the estimation of
$\nu$ from the ansatz in Eq.(\ref{E.ScalingAnsatz}), by looking
for the best exponent for which the data collapse onto a single
curve. A sample of our results is shown in the lower-right inset
of Fig.\ref{F.magz}. From such an analysis, we obtain $\nu=1$ and,
therefore, $\tilde{\beta}=\frac{1}{2}$, a result that fully agrees
with the critical exponent for $\langle \hat{\sigma}_1^z \rangle$
derived in  Sec.~\ref{Ss.TM} in the thermodynamic limit.

\begin{figure}[ht]
   \includegraphics[width=\linewidth]{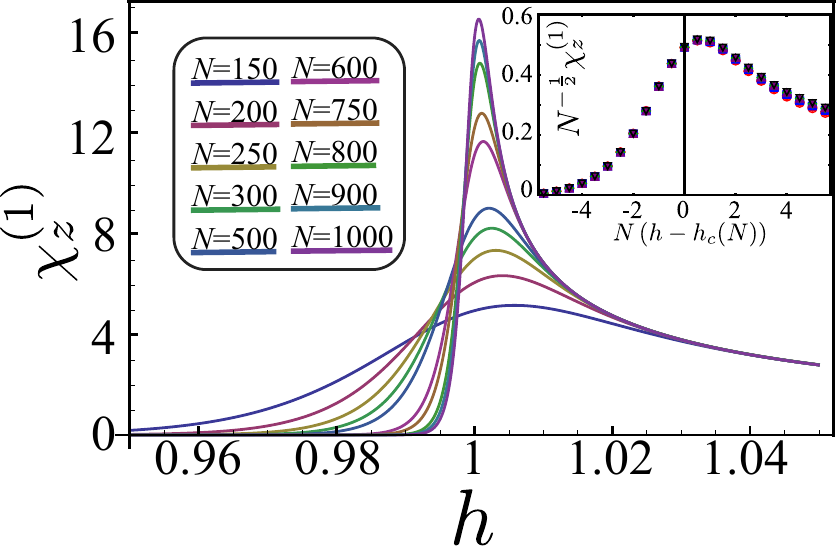} \\
\vspace{0.3cm}
    \includegraphics[width=0.49\linewidth]{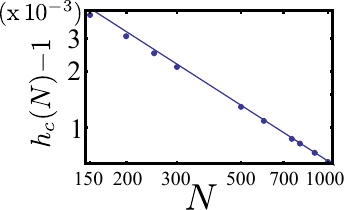}
   \includegraphics[width=0.49\linewidth]{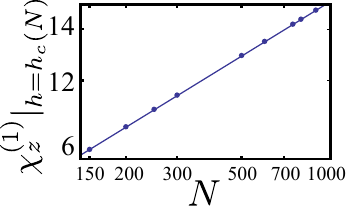}

\caption{(color online): \\
{\it Upper plot} - Impurity susceptibility $\chi_z^{(1)}$  for
different system sizes, in the vicinity of $h=1$, where a
diverging behavior is clearly displayed. {\it inset} -
Data collapse for the impurity transverse susceptibility obtained
with scaling exponents $\nu=1$ and $\alpha=\frac{1}{2}$. {\it Lower left plot} -
Log-log plot of the pseudocritical point $h_c(N)$ vs. $N$,
obtained from the position of the maximum of $\chi_z^{(1)}$, which
allows to extract the shift exponent $\lambda$.   {\it Lower right plot} -
Plot of the maximum of $\chi_z^{(1)}$ vs. $N$ which shows the logarithmic nature of the divergence reported in the upper plot. }
\label{F.susg1}
\end{figure}

Next, we move on to discuss   the finite size scaling of the
impurity transverse susceptibility. As shown in the main plot of
Fig.\ref{F.susg1}, $\chi_z^{(1)}$ diverges for $h \rightarrow 1$,
in the thermodynamic limit. From the corresponding plots, we can
fit the shift exponent $\lambda$ ruling the approach to $h_c$ of
the pseudocritical value $h_c'$ via the relation $| h'_c-h_c|
\propto N^{-\lambda}$. Our results are consistent with the value
$\lambda=1$ (upper right inset of Fig.\ref{F.susg1}). Finally,
from  the relation $\lambda=\frac{1}{\nu}$, we again obtain
$\nu=1$. Using  the relation $\chi_z^{(1)} |_{h=h'_c}\sim
N^{\frac{\alpha}{\nu}}$, we obtain for the ratio
${\frac{\alpha}{\nu}}$ a value compatible with $\frac{1}{2}$
(lower right inset of Fig.\ref{F.susg1}). Given the above results,
we can now verify the consistency of the data collapse of
$\chi_z^{(1)}$ with the finite size scaling ansatz
\begin{equation}
\label{E.chizFSS}
\chi_z^{(1)} (h)=N^{\frac{\alpha}{\nu}}\tilde{\chi}\left(N^{\frac{1}{\nu}}(h-h_c')\right)~.
\end{equation}
\noindent In the upper left inset of Fig.\ref{F.susg1}, we
actually see  such a data collapse for the predetermined values
$\nu=1$ and for $\alpha=\frac{1}{2}$. Finally, in
addition to the critical exponents for $\chi_z^{(1)} (h)$, and in analogy with the analysis for $\langle \hat{\sigma}_1^z \rangle$, we can
also analytically determine the corresponding scaling. Indeed,
from Eq.(\ref{E.Sz}),  we obtain
\begin{align}
\label{E.susZ}
\chi_z^{(1)}=\frac{d \langle \hat{\sigma}_1^z \rangle}{dh}=
\frac{h^{2\left(N-1\right)}\left(N(1-h^2)+h^{2N}-1\right)}{2\left(h^{2N}-1\right)^2
\sqrt{\frac{h^{2N}(h^2-1)}{h^{2N}-1}}}~.
\end{align}
\noindent
Expanding Eq.(\ref{E.susZ})  for $h\simeq 1$ and for $N\gg 1$, we eventually find
\begin{align}
\label{E.susZ_sca}
\chi_z^{(1)}\sim N^{\frac{1}{2}}\left(\frac{1}{4}
+\left(\frac{N}{12}-1 \right)\frac{h-1}{2} \right)~,
\end{align}
\noindent Eq.(\ref{E.susZ_sca}) states that   the critical
exponent for the impurity transverse susceptibility is
$\frac{1}{2}$, as well. From Eq.(\ref{E.susZ}),  we can also
determine  the position of the maximum, obtaining, once again, the
shift of the pseudocritical  point as $h_c'-1\simeq
\frac{2}{3}N^{-1}$. This indicates that the shift exponent is
$\lambda=1=\frac{1}{\nu}$, in full agreement with previous
theoretical~\cite{domb1983phase} and
numerical~\cite{1742-5468-2008-06-P06004} predictions.

To summarize, so far we have performed a finite size scaling analysis of the
impurity related observables close to the quantum phase transition, obtaining the
following critical exponents: $\nu{=}1$,
$\tilde{\beta}{=}\frac{1}{2}$, and $\alpha=\frac{1}{2}$. It is important
to stress once more  that the condition $\mu=0$ is crucial, in
order to witness the quantum phase transition. Finally, let us remark that  the
susceptivity of the homogenous Ising model, corresponding to the
specific heat in the 2D classical Ising, has a logarithmic
divergence, $\chi_z |_{h=h_m}\sim  \log N$, and the critical
exponent is $\alpha=0$. As a consequence, the impurity
susceptibility signals a qualitatively different phenomenon
occurring at the system's edge. We now move on, to consider the
finite size scaling of the longitudinal  magnetization $ \langle \hat{\sigma}_1^x
\rangle$.

\begin{figure}[ht]
        \begin{center}
     \includegraphics[width=.49\linewidth]{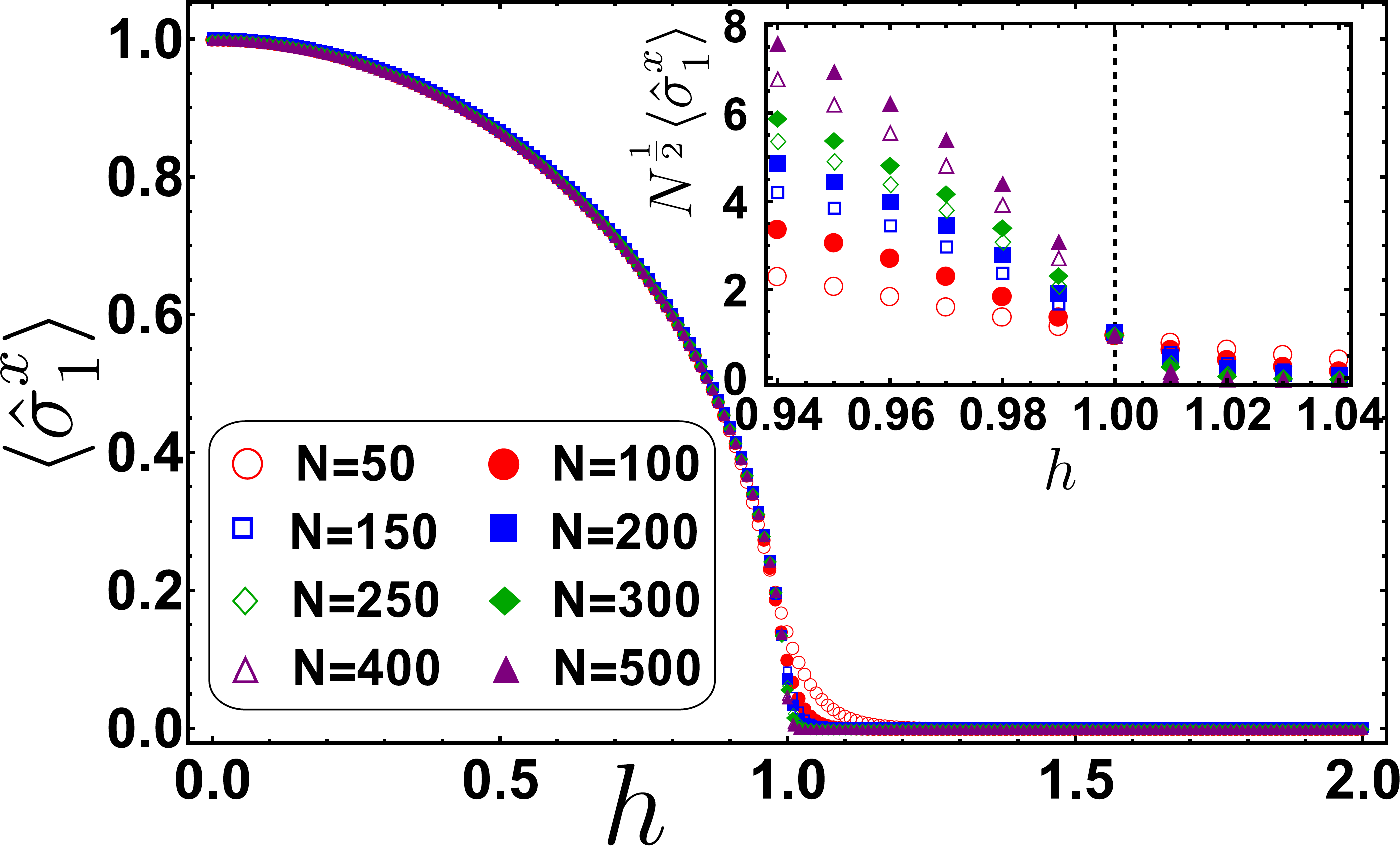}
        \includegraphics[width=.49\linewidth]{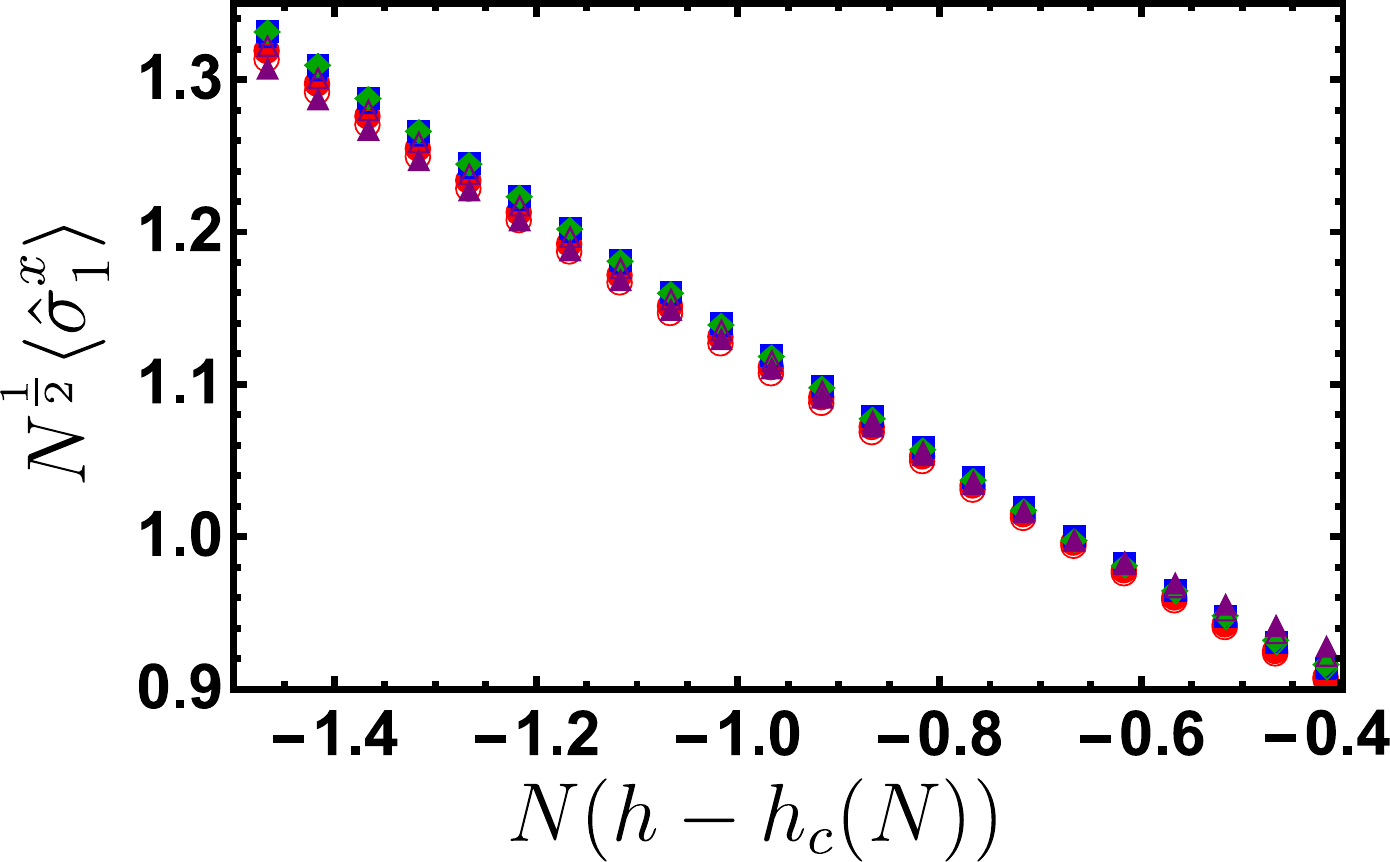}
\caption{(color online): {\it Left plot} - Longitudinal impurity
magnetization, $\langle \hat{\sigma}_1^x \rangle$, as a function
of $h$ for the homogenous finite-size spin chain ($\mu=1$),
evaluated according to Eq.(\ref{E.Xfact}). {\it inset}
- Zoom around the critical point of $N^\frac{\beta}{\nu} \:
\langle \hat{\sigma}_1^x \rangle$ as a function of $h$, with
$\beta = \frac{1}{2} , \nu = 1$. {\it Right plot} -  Data
collapse according to the finite size scaling relation in
Eq.(\ref{E.ScalingAnsatzXhomo}) with $\beta = \frac{1}{2} , \nu =
1$.  }
        \label{F.XXall}
        \end{center}
\end{figure}

\begin{figure}[h]
        \begin{center}
        \includegraphics[width=\linewidth]{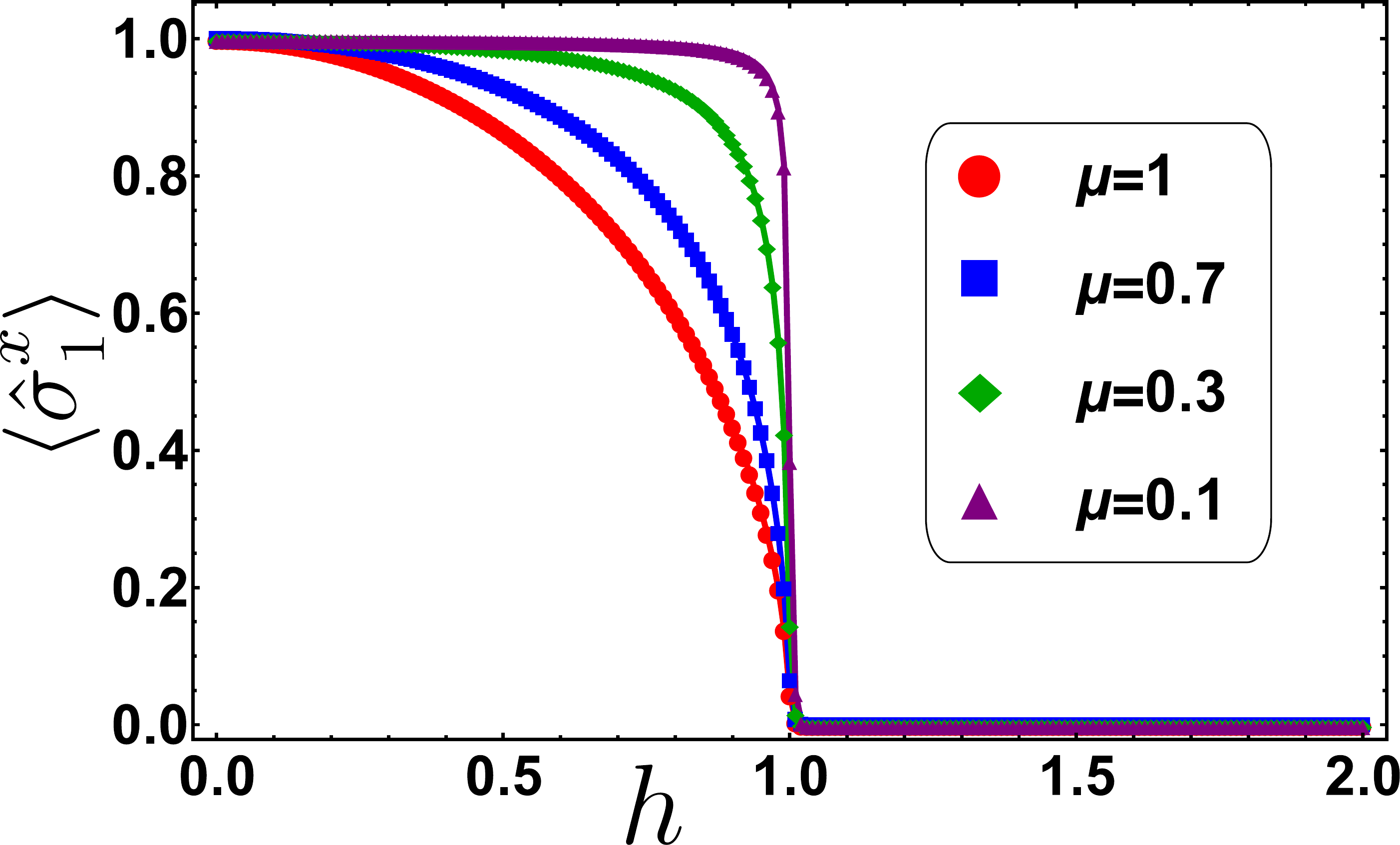}
\caption{ (color online): Longitudinal impurity magnetization
$\average{\hat{\sigma}_1^x}$ for $N = 500$ at  different values of
$\mu$. }
        \label{F.Ximpall}
        \end{center}
\end{figure}

First of all, in order to recover an (approximate) analytic counterpart of the numerical results
we discuss in the following, we have to properly generalize Eq.(\ref{E.sigmax}) to a finite $N$.
Using again the wavefunctions listed in Table \ref{T.Sol}, it is not difficult to show that, for
finite $N$ and for $h \to 1^{- }$, one obtains
\begin{align}\label{fsscaling}
\langle \hat{\sigma}_1^x \rangle &= \sqrt{\frac{1 - h^2}{1 + ( \mu^2 - 1 ) h^2}}\\
&\left\{ 1 + \frac{\mu^2 ( 1 - h^2 ) }{1 + ( \mu^2 - 1) h^2} N h^{2N} + {\cal O} ( N h^{2 N} ) \right\}\nonumber
~.
\end{align}
\noindent
 \noindent
As a preliminary check, in Fig.\ref{F.XXall} we show the finite size scaling data collapse for $\langle \hat{\sigma}_1^x \rangle$ as
a function of $h$ in the absence of the impurity ($\mu = 1$). From the main plot in Fig.\ref{F.XXall},
as $N$ gets large, we recover the expected scaling of $\langle \hat{\sigma}_1^x \rangle$ as $ ( 1 - h )^\frac{1}{2}$ for
$h < 1$, as from Eq.(\ref{E.sigmax}) for $\mu = 1$, while $\langle \hat{\sigma}_1^x \rangle$  experiences a steep decrease to zero for $h > 1$
(disordered phase). To actually check finite size scaling, in the bottom-left inset of Fig.\ref{F.XXall}, we plot $N^\frac{\beta}{\nu} \: \langle \hat{\sigma}_1^x \rangle$,
for various values of $N$ (see figure caption for details). We find the expected intersection between all the curves
drawn for various values of $N$ once we set $\beta = \frac{1}{2} , \nu = 1$, consistently with Eq.(\ref{fsscaling}) taken
for $\mu = 1$ and for $h \to 1^-$, as well as with the finite size scaling hypothesis for
a homogeneous chain. In fact, our result is further corroborated by the plot in the top-right inset of Fig.\ref{F.XXall},
where we draw $N^\frac{\beta}{\nu} \: \langle \hat{\sigma}_1^x \rangle$ for  $\beta = \frac{1}{2} , \nu = 1$, as a function of
the rescaled variable $N^\frac{1}{\nu} ( h - h_c ( N ))$, finding an excellent data collapse, in agreement with
finite size scaling analysis based on  the scaling formula for $\langle \hat{\sigma}_1^x \rangle$, given by
\begin{equation}
 \langle \hat{\sigma}_1^x \rangle =N^{-\frac{\beta}{\nu}}g\left(\left(h-h_c\right)N^{\frac{1}{\nu}}\right)
 \;\;\;\; ,
 \label{E.ScalingAnsatzXhomo}
\end{equation}
\noindent
with $\beta = \frac{1}{2} , \nu = 1$.
For  $\mu \neq 1$, the behavior of the longitudinal magnetization
changes substantially. Indeed, from Fig.\ref{F.Ximpall},
Eq.(\ref{E.ScalingAnsatzXhomo}) appears to be definitely violated
if $\mu < 1$. On the other hand, as soon as $N$ gets large enough,
Eq.(\ref{E.sigmax}), fits well the numerical data at any $\mu$. A
possible explanation of such a finite size scaling violation can
be recovered by combining the results for the phase diagram in
Fig.\ref{Figdiscr} with the finite-$N$ approximate formula for
$\langle \hat{\sigma}_1^x \rangle$ in Eq.(\ref{fsscaling}). From
Fig.\ref{Figdiscr}, one sees that, on varying $h$ from left to
right at a fixed $\mu$,  one actually encounters two phase
transitions. A first one, corresponding to the closure of the bulk
Ising ordered phase at $h \sim 1$, and a second one, corresponding
to the appearance of the mode $\Lambda_2$ at $ | h ( 1 - \mu )|
\geq 1$. The former (ferromagnetic-paramagnetic) quantum phase
transition is a bulk phenomenon; that is, it affects the
longitudinal magnetization at any point of the open Ising chain,
though with different exponents in the bulk and at the boundary.
The latter transition, instead, has to be rather regarded as an
impurity phase transition, not accompanied by a change in the bulk
phase of the system. Nevertheless, in analogy with the emergence
of a dynamically generated Kondo length \cite{affleck08}, a length
scale $\xi_\mu$ different from $\xi (h)$ seems to appear near this
second transition, with eventually $\xi_\mu \to \infty$ right at
the critical point. An approximate explicit formula for $\xi_\mu$
valid for $h \sim 1$ can be inferred from Eq.(\ref{fsscaling}),
which can be rewritten as
\begin{equation}
 \langle \hat{\sigma}_1^x \rangle = \frac{1}{\mu} \: \sqrt{\frac{1}{\xi ( h ) + \xi_\mu}} \: \left\{ 1 + \frac{N}{\xi ( h )  + \xi_\mu} \:
 e^{ - \frac{N}{\xi ( h ) } } \right\}
 \:\:\:\: ,
 \label{fscaling.2}
\end{equation}
\noindent with $\xi_\mu = \mu^2 / | 1 - \mu^2 | $.
Eq.(\ref{fscaling.2}) provides a possible explanation of the
scaling violation for $\mu \in (0,1)$. Indeed, when comparing it
with the generic finite size scaling ansatz formula in
Eq.(\ref{E.ScalingAnsatzXhomo}), we see that, in order for the two
of them to be consistent with each other,
Eq.(\ref{E.ScalingAnsatzXhomo}), which can be recast in the form $
\langle \hat{\sigma}_1^x \rangle =N^{-\frac{\beta}{\nu}} \:
\tilde{g} \left( \frac{N}{\xi ( h ) } \right)$, must be
generalized to an expression of the general form
\begin{equation}
  \langle \hat{\sigma}_1^x \rangle =N^{-\frac{\beta}{\nu}} \: \phi \left[ \frac{N}{\xi ( h )  } , \frac{N}{\xi_\mu} \right]
  \:\:\:\: .
  \label{fscaling.3}
\end{equation}
\noindent While it would be definitely of interest to corroborate
the two-length scenario with, for instance, a detailed analytical
derivation of the formula for $\xi_\mu ( h )$ at a generic value
of $h$, this would require a careful analytical study of the transverse field Ising model
in the presence of a boundary impurity, which goes beyond the
scope of this paper, focused on the discussion of the
finite size scaling in the transverse field Ising model on a finite chain. Therefore, while we plan to
address in detail this issue in a forthcoming publication, here we
limit ourselves to a few additional observations on the two-length
generalized scaling formula in Eq.(\ref{fscaling.3}). First of
all, we note that, as $\mu \to 1$ (that is, when going back to
the homogeneous chain case), one gets $\xi_\mu \to \infty$, which
implies that, in this limit, Eq.(\ref{fscaling.3}) reduces back to
the standard, single-length finite size scaling formula in
Eq.(\ref{E.ScalingAnsatzXhomo}), as it must be.

\begin{figure}[ht]
        \begin{center}
        \includegraphics[width=\linewidth]{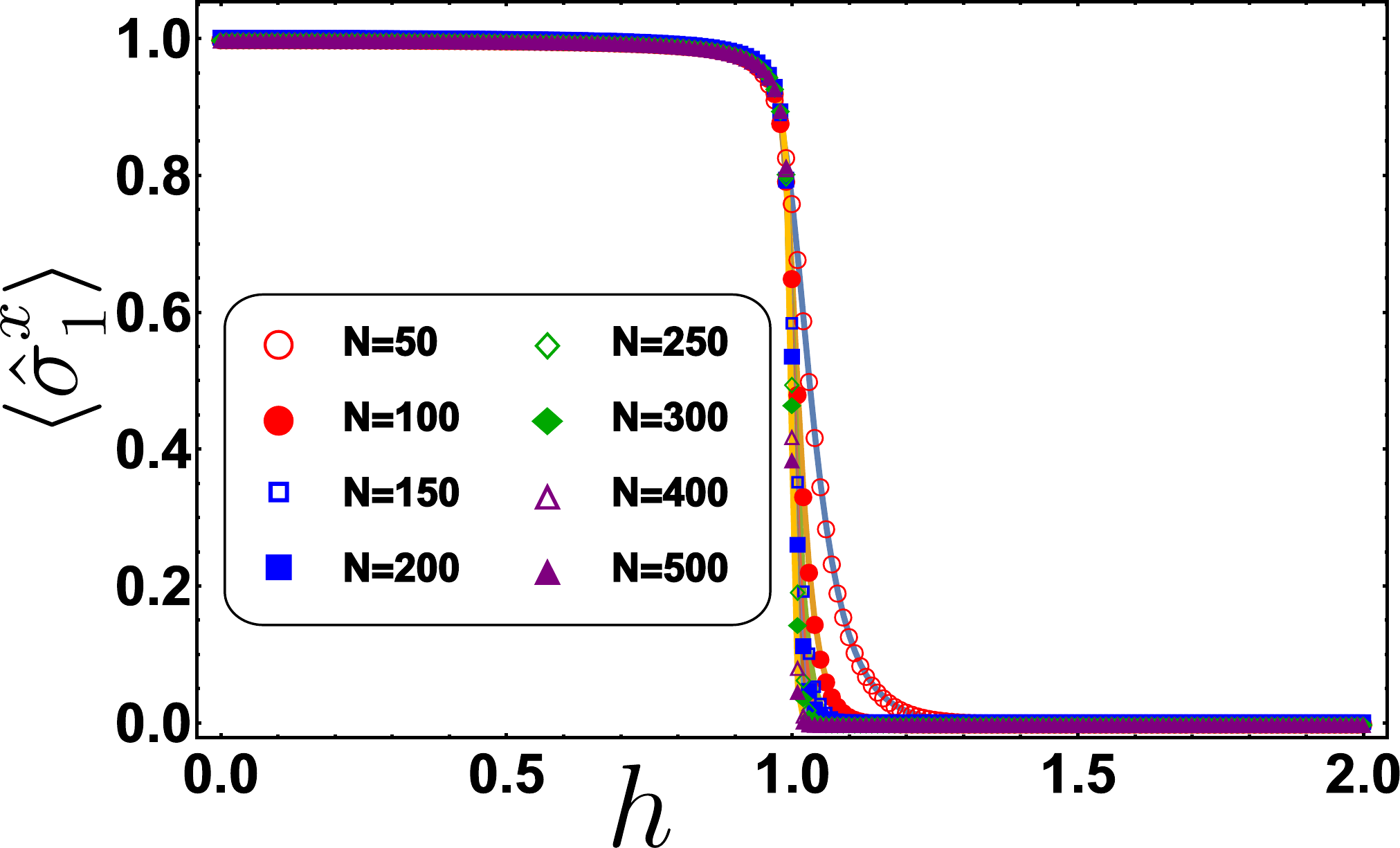}
\caption{(color online):  Longitudinal impurity magnetization
$\average{\hat{\sigma}_1^x}$ for different values of $N$ evaluated
for $\mu=0.1$.  }
        \label{F.Ximp_col}
        \end{center}
\end{figure}

In the complementary limiting case, $\mu \to 0$, we have $\xi_\mu
\to 0$; then, looking at Eq.(\ref{fscaling.3}), one would again
expect a finite size scaling formula such as in  Eq.(\ref{E.ScalingAnsatzXhomo}).
In fact, this is quite a peculiar scaling limit, in that, as $\mu
\to 0$, one finds that $\hat{\sigma}_1^x $ becomes an exactly
conserved quantity, even at finite $N$. In Fig.\ref{F.Ximp_col},
we plot $\langle \hat{\sigma}_1^x \rangle$ as a function of $h$
for a rather small value of $\mu$ (the $\mu = 0$ limit was hard to
recover in the numerical calculations, due to the increasing lack
of numerical precision for $\mu < 0.1$). Apparently,
Fig.\ref{F.Ximp_col}  is consistent with the previous discussion,
since $\langle \hat{\sigma}_1^x \rangle$ remains basically
constant and finite, for $h < 1$, while it suddenly jumps to 0, as
soon as $h$ crosses the quantum phase transition point. Looking back at
Eq.(\ref{E.ScalingAnsatzXhomo}), one may actually  state that this
behavior corresponds to an effective $\beta = 0$, which appears to
be consistent with $\hat{\sigma}_1^x$ being an exactly conserved
quantity for $\mu = 0$, at any value of $N$.
\noindent
\noindent
\noindent
As a one-sentence summary of this section, one can state that,
while the longitudinal magnetization $\langle \hat{\sigma}_1^x
\rangle$ exhibits a neat finite size scaling with $\beta =
\frac{1}{2} , \nu = 0$ at $\mu = 1$, with increasing deviation
from this behavior as $\mu$ moves from $1$ to $0$, the transverse
magnetization  $\langle \hat{\sigma}_1^z \rangle$ shows a sort of
fully complementary behavior: no scaling at any $\mu > 0$, with an
emerging scaling behavior recovered at $\mu = 0$.

\section{Finite size scaling in the $XY$ model}
\label{S.XY}

In this section, we test the universality of the impurity magnetic
observables scaling properties by extending the previous analysis
to the  $XY$-model, which belongs to the same universality class
as the Ising model. To this end,  we consider the following model
Hamiltonian, describing a side impurity in the $XY$-chain,
\begin{align}
\label{E.HXYspin}
\hat{H}^{XY}_ {\mu }&{=}{-}J
\left[\sum_{n{=}1}^{N{-}1}  \{ \left(1{+}\gamma_n\right) \hat{\sigma}_n^x \hat{\sigma}_{n{+}1}^x{+}\left(1{-}\gamma_n\right)
\hat{\sigma}_n^y \hat{\sigma}_{n{+}1}^y \}\right.\nonumber\\
&{+}\left. \sum_{ n {=} 1}^N h_n \hat{\sigma}_n^z\right]~,
\end{align}
\noindent
with the side impurity at $n=1$ defined by setting
\begin{eqnarray}
 \gamma_n &=& \gamma ( 1 - \delta_{n1} ) + \gamma_1 \delta_{1n} \nonumber \\
 h_n &=& h ( 1 - \delta_{ n1} ) + \mu h \delta_{n1}
 \:\:\:\:.
 \label{parameters.xy}
\end{eqnarray}
\noindent
This choice corresponds to both the magnetic field and
the anisotropy parameter staying homogeneous on the bulk, with
values $h$ and $\gamma$, but taking different values at the edge
of the chain, where they become $\mu h$ and $\gamma_1$,
respectively.

With the bulk anisotropy parameter $\gamma\in\left(0,1\right]$,
the $XY$ model undergoes a quantum phase transition at a critical
value of $h=h_c$, which falls into the same universality class as
the transverse field Ising model. Our aim here is to investigate
whether the bulk equivalence extends to the impurity finite size
scaling, as well.

\begin{figure}[ht]
        \begin{center}
        \includegraphics[width=\linewidth]{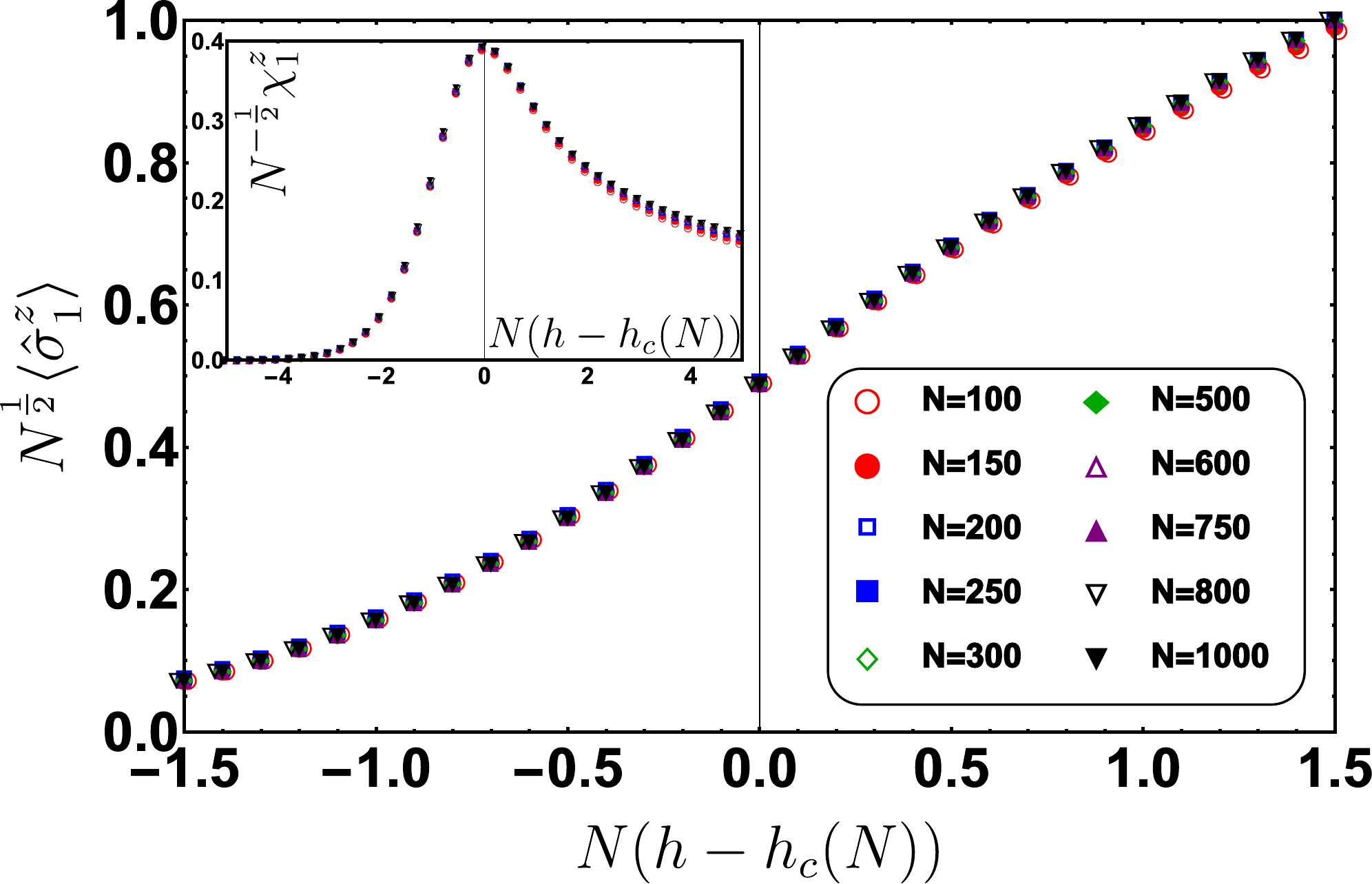}
\caption{(color online): {\it Main plot} -  Data collapse of the
transverse impurity magnetization $\langle \hat{\sigma}_1^z
\rangle$ for an $XY$ model, Eq.(\ref{E.HXYspin}),  with $\gamma=0.5$ and $\gamma_1=1$.
{\it Inset} -  Transverse local spin susceptibility for the same value of $\gamma$.
The critical exponents take the same values as in the plots in
Fig.\ref{F.magz} and in Fig.\ref{F.susg1}, respectively. }
        \label{F.FSSgamma05mag}
        \end{center}
\end{figure}

An analytical solution of the impurity  model for $\gamma\neq 1$
analogous to that provided in Ref.[\onlinecite{apollaro}] for the
Ising model is lacking. Nevertheless, one may readily check that,
as $\mu \to 0$, the parity operator $\hat{P}$ is again a good
symmetry of the model, that is, $ [ \hat{H}_{ XY ; \mu = 0  } ,
\hat{P} ] = 0$, provided that  $\gamma_1=\pm 1$ (which means
 a Ising-like coupling, either along the
$X$ or  the $Y$ direction). One can then analytically verify that
both possibilities imply the presence of a zero-energy eigenvalue,
in analogy to what happens in the transverse field Ising model  in
Eq.(\ref{E.HIsing}). It is hence reasonable to expect that, also
in this case, one recovers the behavior of a classical edge
impurity, just as in the Ising model. Indeed, we obtain, for the
whole Ising universality class, $\gamma \in \left(0,1\right]$, the
same finite size scaling exponents as in the  transverse field
Ising model; that is, $\nu=1$, $\tilde{\beta}=\frac{1}{2}$, and
$\alpha=\frac{1}{2}$. To highlight this result, in
Fig.\ref{F.FSSgamma05mag}, we show  the  data collapse of the
transverse impurity magnetization and susceptibility, according to
Eqs.(\ref{E.ScalingAnsatz}) and(\ref{E.chizFSS}) for the
$XY$-model with bulk  anisotropy $\gamma=0.5$.
 \noindent

\section{The classical-to-quantum correspondence in the Ising model with an impurity}
\label{cobc}  The one-dimensional quantum Ising model can be
mapped onto the two-dimensional classical model, as a special case
of the general correspondence between critical phenomena in
$d$-dimensional quantum statistical models and those in
$d{+}1$-dimensional classical statistical models. Such a
correspondence has been largely discussed and reviewed in the
literature, particularly in the context of lattice gauge theories
and classical, as well as quantum, lattice spin models
\cite{kogut_79,fradkin_78}.  The classical 2-dimensional Ising
model is one of the best known classical statistical systems,
especially (but not only) in view of the remarkable Onsager's
exact solution of the model (see, for instance,
Ref.[\onlinecite{mccoy}] for an extensive review on the topic).
According to this correspondence, one finds, e.g., that the
transverse field Ising model with periodic boundary conditions is
related to a $d=2$ classical Ising model on a torus
\cite{LIEB1961407}, while the homogeneous transverse field Ising
model with open boundary conditions corresponds to a finite-size
cylinder whose boundary circles map onto the (quantum) endpoints
of the chain \cite{mccoy,peschel_1}. One may generalize the
mapping to the case of models with generic, position-dependent
parameters as outlined in Ref.[\onlinecite{peschel_2}]. Following
\cite{peschel_2}, in this section we construct the mapping between
the inhomogeneous transverse field Ising model with an impurity at
its left-hand side endpoint and the corresponding $2d$-classical
spin model.

Following the approach discussed in  Appendix B of
Ref.[\onlinecite{peschel_2}], we choose as a reference
$2d$-classical system the planar Ising model with inhomogenous
couplings, described by the Hamiltonian

\begin{widetext}
\begin{equation}
{\cal E} [ \{ \sigma_{ n , m } \} ] =
- \sum_{ n = 1}^{M_x} \sum_{ m = 1}^{ M_y } K_y ( n ) \sigma_{ n , m }
\sigma_{ n , m+1}
- \sum_{ n = 1}^{M_x - 1} \sum_{ m = 1}^{ M_y } K_x ( n ) \sigma_{ n , m }
\sigma_{ n + 1 , m}
\:\:\:\: ,
\label{2d.1b}
\end{equation}
\end{widetext}
\noindent with the couplings  $K_x $ and $K_y$ generically
depending on $n$ and with $\sigma_{n , m}$ being a classical
(binary) spin variable at site $n , m$ of the lattice (which we
shall identify with the $x$-component of the quantum spin when
mapping to a $1d$ inhomogeneous transverse field Ising model) and
with periodic boundary conditions assumed in the $y$-direction,
that is, $\sigma_{ n , m + M_y } = \sigma_{n , m }$, $\forall n =
1 , \ldots , M_x$ (so that, pictorially, the lattice may be
thought as wrapped around a cylinder along the $y$-direction). At
temperature $T= \beta^{-1}$, one  obtains the classical partition
function for the model in Eq.(\ref{2d.1b}) as
\begin{widetext}
\begin{equation}
{\cal Z} [ \{ K_x , K_y \} ; T ] = \sum_{ \{ \sigma_{ n , m }  =
\pm 1 \} } \: \exp\left[{ \beta [  \sum_{ n = 1}^{M_x} \sum_{ m = 1}^{ M_y
} K_y (n ) \sigma_{n ,m } \sigma_{ n , m+1} +   \sum_{ n = 1}^{M_x
- 1} \sum_{ m = 1}^{ M_y } K_x ( n ) \sigma_{ n , m } \sigma_{ n+
1 , m} ] }\right] \:\: . \label{2d.1.5}
\end{equation}
\end{widetext}
\noindent To complete the mapping onto the $1d$-quantum Ising
model, one has to  to rewrite Eq.(\ref{2d.1.5}) as the partition
function of a 1-dimensional transverse field Ising model with
non-homogeneous parameters. This requires constructing   an
appropriate transfer matrix ${\bf T}$,  whose matrix elements can
be regarded as the partition function for two subsequent rows (say
$m$ and $m+1$)  with assigned spin configurations $ \{ | \{
\sigma_{ n, m } \}_{n = 1, \ldots M_x} \rangle_m$ and $ \{ |  \{
\sigma_{ n , m+1 }' \}_{n = 1, \ldots M_x} \rangle_{m+1} \}$, so
that

\begin{equation}
[ T]_{ \{ \sigma_{n}' \} , \{ \sigma_{ n } \}} =   ~_{m+1}
\langle \sigma_{n}' |   {\bf T}  | \sigma_{n} \rangle_m
\:\:\:\: .
\label{2d.2}
\end{equation}
\noindent
In the low-$T$ (large-$\beta$)-limit one may explicitly write ${\bf T}$ as
\cite{peschel_2}

\begin{equation}
{\bf T} = \exp [ - \hat{H}_{1d } ]
\;\;\;\; ,
\label{corr.1}
\end{equation}
\noindent
with $\hat{H}_{1d }$ being the Hamiltonian of the
inhomogenous quantum Ising chain  given by (apart for an
irrelevant over-all constant term)
\begin{equation}
\hat{H}_{1d }= -
\sum_{ n = 1}^{M_x} h_n \hat{\sigma}_n^z -    \sum_{ n = 1}^{M_x - 1 }
J_n \hat{\sigma}_n^x \hat{\sigma}_{ n + 1}^x
\;\;\;\; ,
\label{corr.2}
\end{equation}
\noindent
with

\begin{eqnarray}
J_n &=& \beta K_x  ( n ) \nonumber \\
e^{-2 h_n} &=& \tanh  [ \beta K_y ( n) ]
\:\:\:\:\: ,
\label{corr.3}
\end{eqnarray}
\noindent
and $\hat{\sigma}_n^{x,z}$ being quantum spin operators on a
$1d$-lattice.

Eq.(\ref{corr.3}) gives the classical correspondent of the quantum
Ising chain with an impurity at its leftmost site. In fact, one
has to assume that Eq.(\ref{corr.3}) holds with $h_n$ and $K_y (
n)$ independent of $n$ at any $n$ but for $n=1$. Basically, the
classical model can be regarded as a cylinder that, as $N \to
\infty$, goes all the way to $\infty $ towards the positive-$x$
axis. The cylinder is cut in correspondence of the first site of
the quantum chain. The smaller is $h_1$, the stronger is  the
effective Ising coupling $\beta K_y ( 1 )$ on the circumference
running around the cut. Eventually, the limit of classical
impurity in the chain ($h_1 \to 0$) corresponds to the limit
$\beta K_y ( 1 ) \to \infty$, in which, consistently with what one
expects from the classical-to-quantum correspondence, all the
thermal fluctuations are suppressed in the classical Ising chain
lying over the circle at $n=1$.

\section{Concluding remarks}
\label{S.Concl}

In this paper, we have studied finite size scaling for an impurity
residing at the boundary (first site) of an otherwise homogeneous
quantum Ising chain in a transverse magnetic field $h$, modelled
by rescaling $h$ on the first site by $h \to h_1= \mu h$. To
highlight finite size scaling at the impurity, we computed in a
finite-size chain (with $N$ sites) both the transverse and the
longitudinal magnetization at the first site of the chain. This
allowed us to single out the limits $\mu \to 1$ and $\mu \to 0$ as
the two regimes in which ``standard'' finite size scaling is
recovered, though in complementary fashions. Indeed, while in the
homogenous chain limit ($\mu \to 1$) $\langle \hat{\sigma}_1^x
\rangle$ exhibits finite size scaling with $N^{ -
\frac{\beta}{\nu}}$, with $\beta = \frac{1}{2}$, and $\nu = 1$,
with $\langle \hat{\sigma}_1^z \rangle$ showing no finite size
scaling at all, in the complementary ``classical impurity'' limit
($\mu \to 0$), full finite size scaling is recovered for the
transverse magnetization, $\langle \hat{\sigma}_1^z \rangle$, with
the same exponents reported above. On the other hand, the
longitudinal magnetization, $\langle \hat{\sigma}_1^x \rangle$,
does not show scaling at $\mu=0$. This nicely corresponds the fact
that $\hat{\sigma}_1^x$ is an exactly conserved quantity in the
finite chain, at $\mu = 0$. In this limit, on extending the finite
size scaling analysis to the impurity transverse susceptibility
$\chi_z^{(1)} ( h) $, we have found again  results consistent with
the finite size scaling ansatz and, in addition, have been able to
recover the shift exponent governing the location of the quantum
pseudocritical point at finite $N$.

Our analysis, thus, extends the finite size scaling behavior of
the homogenous transverse field Ising model
\cite{kim,0305-4470-18-1-006} with open boundary conditions to the
case in which an impurity is present at one endpoint of the chain
by showing that, as  $\mu = 0$, the transverse surface
magnetization, $\langle \hat{\sigma}_1^z \rangle$, works as a sort
of ``disorder'' parameter for the bulk (being nonzero in the
disordered phase), with the same exponents characterizing the
finite size scaling of $\langle \hat{\sigma}_1^x \rangle$ for the
homogeneous model ($\mu = 1$). As mentioned in the discussion of
Sec.~\ref{S.FSS}, the intermediate situation $0< \mu <1$ is much
less clear, in that the scaling ansatz of
Eq.(\ref{E.ScalingAnsatzXhomo}) appears not to reproduce the
finite size behavior of $\langle \hat{\sigma}_1^x \rangle$. In
fact, the smaller $\mu$, the more severe it is the deviation of
$\langle \hat{\sigma}_1^x \rangle$ from the general scaling
formula. As outlined above, a possible explanation, consistent
with the asymptotic expressions for $\langle \hat{\sigma}_1^x
\rangle$ in Eqs.(\ref{E.sigmax},\ref{fsscaling}), relies  on the
emergence of two length scales, associated with the two QPTs the
system goes through, for $\mu \neq 1$. While this appears to be a
suggestive and promising hypothesis, its careful analytical and
numerical verification goes beyond the scope of this work and,
accordingly, we chose to postpone it to a forthcoming publication.

Finally, on employing the standard quantum-to-classical mapping
for our specific impurity model, we have been able to construct
the (two-dimensional) statistical model providing the classical
counterpart of a transverse field Ising model with an impurity at one of its endpoints as a
classical Ising model on a rectangular lattice wrapped on a
half-infinite cylinder, with the links along the first circle
altered, as a function of $\mu$, according to Eq.(\ref{corr.3}).

\vspace{0.5cm}

\section{Ackowledgments}
We thank A.~Nersesyan for useful correspondence and N. Lo Gullo, S. Lorenzo,  A. Gambassi, and F. Franchini for insightful discussions. T.J.G.A.,
G.F., G.M.P. and F.P. acknowledge support from the Collaborative
Project QuProCS (Grant Agreement 641277).

\bibliography{test_rev}

\begin{thebibliography}{60}
\expandafter\ifx\csname natexlab\endcsname\relax\def\natexlab#1{#1}\fi
\expandafter\ifx\csname bibnamefont\endcsname\relax
  \def\bibnamefont#1{#1}\fi
\expandafter\ifx\csname bibfnamefont\endcsname\relax
  \def\bibfnamefont#1{#1}\fi
\expandafter\ifx\csname citenamefont\endcsname\relax
  \def\citenamefont#1{#1}\fi
\expandafter\ifx\csname url\endcsname\relax
  \def\url#1{\texttt{#1}}\fi
\expandafter\ifx\csname urlprefix\endcsname\relax\def\urlprefix{URL }\fi
\providecommand{\bibinfo}[2]{#2}
\providecommand{\eprint}[2][]{\url{#2}}

\bibitem[{\citenamefont{Sachdev}(2011)}]{sachdev2011}
\bibinfo{author}{\bibfnamefont{S.}~\bibnamefont{Sachdev}},
  \emph{\bibinfo{title}{Quantum phase transitions}}
  (\bibinfo{publisher}{Cambridge University Press},
  \bibinfo{address}{Cambridge}, \bibinfo{year}{2011}), \bibinfo{edition}{2nd}
  ed.

\bibitem[{\citenamefont{Jaeger}(1998)}]{Jaeger1998}
\bibinfo{author}{\bibfnamefont{G.}~\bibnamefont{Jaeger}},
  \bibinfo{journal}{Archive for History of Exact Sciences}
  \textbf{\bibinfo{volume}{53}}, \bibinfo{pages}{51} (\bibinfo{year}{1998}).

\bibitem[{\citenamefont{Griffiths}(1970)}]{griffiths}
\bibinfo{author}{\bibfnamefont{R.~B.} \bibnamefont{Griffiths}},
  \bibinfo{journal}{Phys. Rev. Lett.} \textbf{\bibinfo{volume}{24}},
  \bibinfo{pages}{1479} (\bibinfo{year}{1970}).

\bibitem[{\citenamefont{Kadanoff}(1966)}]{kadanoff}
\bibinfo{author}{\bibfnamefont{L.~P.} \bibnamefont{Kadanoff}},
  \bibinfo{journal}{Physics} \textbf{\bibinfo{volume}{2}}, \bibinfo{pages}{263}
  (\bibinfo{year}{1966}).

\bibitem[{\citenamefont{Zinn-Justin}(2012)}]{zinn_justin}
\bibinfo{author}{\bibfnamefont{J.}~\bibnamefont{Zinn-Justin}},
  \emph{\bibinfo{title}{Quantum Field Theory and Critical Phenomena}}
  (\bibinfo{publisher}{Clarendon Press, Oxford}, \bibinfo{year}{2012}).

\bibitem[{\citenamefont{Wilson}(1975)}]{wilson}
\bibinfo{author}{\bibfnamefont{K.~G.} \bibnamefont{Wilson}},
  \bibinfo{journal}{Rev. Mod. Phys.} \textbf{\bibinfo{volume}{47}},
  \bibinfo{pages}{773} (\bibinfo{year}{1975}).

\bibitem[{\citenamefont{{Lorenzo} et~al.}(2017)\citenamefont{{Lorenzo},
  {Marino}, {Plastina}, {Palma}, and {Apollaro}}}]{lorenzo_1}
\bibinfo{author}{\bibfnamefont{S.}~\bibnamefont{{Lorenzo}}},
  \bibinfo{author}{\bibfnamefont{J.}~\bibnamefont{{Marino}}},
  \bibinfo{author}{\bibfnamefont{F.}~\bibnamefont{{Plastina}}},
  \bibinfo{author}{\bibfnamefont{G.~M.} \bibnamefont{{Palma}}},
  \bibnamefont{and} \bibinfo{author}{\bibfnamefont{T.~J.~G.}
  \bibnamefont{{Apollaro}}}, \bibinfo{journal}{Scientific Reports}
  \textbf{\bibinfo{volume}{7}}, \bibinfo{eid}{5672} (\bibinfo{year}{2017}).

\bibitem[{\citenamefont{Osterloh et~al.}(2002)\citenamefont{Osterloh, Amico,
  Falci, and Fazio}}]{Osterloh2002}
\bibinfo{author}{\bibfnamefont{A.}~\bibnamefont{Osterloh}},
  \bibinfo{author}{\bibfnamefont{L.}~\bibnamefont{Amico}},
  \bibinfo{author}{\bibfnamefont{G.}~\bibnamefont{Falci}}, \bibnamefont{and}
  \bibinfo{author}{\bibfnamefont{R.}~\bibnamefont{Fazio}},
  \bibinfo{journal}{Nature} \textbf{\bibinfo{volume}{416}},
  \bibinfo{pages}{608} (\bibinfo{year}{2002}).

\bibitem[{\citenamefont{Campbell et~al.}(2013)\citenamefont{Campbell, Mazzola,
  De~Chiara, Apollaro, Plastina, Busch, and
  Paternostro}}]{CampbellMDGAPBPNJP13}
\bibinfo{author}{\bibfnamefont{S.}~\bibnamefont{Campbell}},
  \bibinfo{author}{\bibfnamefont{L.}~\bibnamefont{Mazzola}},
  \bibinfo{author}{\bibfnamefont{G.}~\bibnamefont{De~Chiara}},
  \bibinfo{author}{\bibfnamefont{T.~J.~G.} \bibnamefont{Apollaro}},
  \bibinfo{author}{\bibfnamefont{F.}~\bibnamefont{Plastina}},
  \bibinfo{author}{\bibfnamefont{T.}~\bibnamefont{Busch}}, \bibnamefont{and}
  \bibinfo{author}{\bibfnamefont{M.}~\bibnamefont{Paternostro}},
  \bibinfo{journal}{New Journal of Physics} \textbf{\bibinfo{volume}{15}},
  \bibinfo{pages}{043033} (\bibinfo{year}{2013}).

\bibitem[{\citenamefont{De~Chiara et~al.}(2012)\citenamefont{De~Chiara, Lepori,
  Lewenstein, and Sanpera}}]{PhysRevLett.109.237208}
\bibinfo{author}{\bibfnamefont{G.}~\bibnamefont{De~Chiara}},
  \bibinfo{author}{\bibfnamefont{L.}~\bibnamefont{Lepori}},
  \bibinfo{author}{\bibfnamefont{M.}~\bibnamefont{Lewenstein}},
  \bibnamefont{and} \bibinfo{author}{\bibfnamefont{A.}~\bibnamefont{Sanpera}},
  \bibinfo{journal}{Phys. Rev. Lett.} \textbf{\bibinfo{volume}{109}},
  \bibinfo{pages}{237208} (\bibinfo{year}{2012}).

\bibitem[{\citenamefont{Bayat et~al.}(2016)\citenamefont{Bayat, Apollaro,
  Paganelli, De~Chiara, Johannesson, Bose, and Sodano}}]{PhysRevB.93.201106}
\bibinfo{author}{\bibfnamefont{A.}~\bibnamefont{Bayat}},
  \bibinfo{author}{\bibfnamefont{T.~J.~G.} \bibnamefont{Apollaro}},
  \bibinfo{author}{\bibfnamefont{S.}~\bibnamefont{Paganelli}},
  \bibinfo{author}{\bibfnamefont{G.}~\bibnamefont{De~Chiara}},
  \bibinfo{author}{\bibfnamefont{H.}~\bibnamefont{Johannesson}},
  \bibinfo{author}{\bibfnamefont{S.}~\bibnamefont{Bose}}, \bibnamefont{and}
  \bibinfo{author}{\bibfnamefont{P.}~\bibnamefont{Sodano}},
  \bibinfo{journal}{Phys. Rev. B} \textbf{\bibinfo{volume}{93}},
  \bibinfo{pages}{201106} (\bibinfo{year}{2016}).

\bibitem[{\citenamefont{Affleck}(2004)}]{affleck08}
\bibinfo{author}{\bibfnamefont{I.}~\bibnamefont{Affleck}}, in
  \emph{\bibinfo{booktitle}{Exact Methods in Low-dimensional Statistical
  Physics and Quantum Computing}}, edited by
  \bibinfo{editor}{\bibfnamefont{J.}~\bibnamefont{Jacobsen}},
  \bibinfo{editor}{\bibfnamefont{S.}~\bibnamefont{Ouvry}},
  \bibinfo{editor}{\bibfnamefont{V.}~\bibnamefont{Pasquier}},
  \bibinfo{editor}{\bibfnamefont{D.}~\bibnamefont{Serban}}, \bibnamefont{and}
  \bibinfo{editor}{\bibfnamefont{L.}~\bibnamefont{Cugliandolo}}
  (\bibinfo{publisher}{Oxford University Press}, \bibinfo{address}{Oxford},
  \bibinfo{year}{2004}).

\bibitem[{\citenamefont{Lorenzo et~al.}(2013)\citenamefont{Lorenzo, Apollaro,
  Sindona, and Plastina}}]{qst1}
\bibinfo{author}{\bibfnamefont{S.}~\bibnamefont{Lorenzo}},
  \bibinfo{author}{\bibfnamefont{T.~J.~G.} \bibnamefont{Apollaro}},
  \bibinfo{author}{\bibfnamefont{A.}~\bibnamefont{Sindona}}, \bibnamefont{and}
  \bibinfo{author}{\bibfnamefont{F.}~\bibnamefont{Plastina}},
  \bibinfo{journal}{Phys. Rev. A} \textbf{\bibinfo{volume}{87}},
  \bibinfo{pages}{042313} (\bibinfo{year}{2013}).

\bibitem[{\citenamefont{Lorenzo et~al.}(2015)\citenamefont{Lorenzo, Apollaro,
  Paganelli, Palma, and Plastina}}]{qst2}
\bibinfo{author}{\bibfnamefont{S.}~\bibnamefont{Lorenzo}},
  \bibinfo{author}{\bibfnamefont{T.~J.~G.} \bibnamefont{Apollaro}},
  \bibinfo{author}{\bibfnamefont{S.}~\bibnamefont{Paganelli}},
  \bibinfo{author}{\bibfnamefont{G.~M.} \bibnamefont{Palma}}, \bibnamefont{and}
  \bibinfo{author}{\bibfnamefont{F.}~\bibnamefont{Plastina}},
  \bibinfo{journal}{Phys. Rev. A} \textbf{\bibinfo{volume}{91}},
  \bibinfo{pages}{042321} (\bibinfo{year}{2015}).

\bibitem[{\citenamefont{Kondo}(1964)}]{kondo}
\bibinfo{author}{\bibfnamefont{J.}~\bibnamefont{Kondo}},
  \bibinfo{journal}{Progress of Theoretical Physics}
  \textbf{\bibinfo{volume}{32}}, \bibinfo{pages}{37} (\bibinfo{year}{1964}).

\bibitem[{\citenamefont{Hewson}(1993)}]{hewson}
\bibinfo{author}{\bibfnamefont{A.~C.} \bibnamefont{Hewson}},
  \emph{\bibinfo{title}{The Kondo Effect to Heavy Fermions}}
  (\bibinfo{publisher}{Cambridge University Press}, \bibinfo{year}{1993}).

\bibitem[{\citenamefont{Furusaki and Hikihara}(1998)}]{furusaki98}
\bibinfo{author}{\bibfnamefont{A.}~\bibnamefont{Furusaki}} \bibnamefont{and}
  \bibinfo{author}{\bibfnamefont{T.}~\bibnamefont{Hikihara}},
  \bibinfo{journal}{Phys. Rev. B} \textbf{\bibinfo{volume}{58}},
  \bibinfo{pages}{5529} (\bibinfo{year}{1998}).

\bibitem[{\citenamefont{Laflorencie et~al.}(2008)\citenamefont{Laflorencie,
  S\o{}rensen, and Affleck}}]{sorensen}
\bibinfo{author}{\bibfnamefont{N.}~\bibnamefont{Laflorencie}},
  \bibinfo{author}{\bibfnamefont{E.~S.} \bibnamefont{S\o{}rensen}},
  \bibnamefont{and} \bibinfo{author}{\bibfnamefont{I.}~\bibnamefont{Affleck}},
  \bibinfo{journal}{Journal of Statistical Mechanics: Theory and Experiment}
  \textbf{\bibinfo{volume}{2008}}, \bibinfo{pages}{P02007}
  (\bibinfo{year}{2008}).

\bibitem[{\citenamefont{Kane and Fisher}(1992)}]{kane92}
\bibinfo{author}{\bibfnamefont{C.~L.} \bibnamefont{Kane}} \bibnamefont{and}
  \bibinfo{author}{\bibfnamefont{M.~P.~A.} \bibnamefont{Fisher}},
  \bibinfo{journal}{Phys. Rev. B} \textbf{\bibinfo{volume}{46}},
  \bibinfo{pages}{15233} (\bibinfo{year}{1992}).

\bibitem[{\citenamefont{Giuliano and Sodano}(2010)}]{giuso0}
\bibinfo{author}{\bibfnamefont{D.}~\bibnamefont{Giuliano}} \bibnamefont{and}
  \bibinfo{author}{\bibfnamefont{P.}~\bibnamefont{Sodano}},
  \bibinfo{journal}{Nuclear Physics B} \textbf{\bibinfo{volume}{837}},
  \bibinfo{pages}{153 } (\bibinfo{year}{2010}).

\bibitem[{\citenamefont{Giuliano and Sodano}(2009)}]{giuso2}
\bibinfo{author}{\bibfnamefont{D.}~\bibnamefont{Giuliano}} \bibnamefont{and}
  \bibinfo{author}{\bibfnamefont{P.}~\bibnamefont{Sodano}},
  \bibinfo{journal}{EPL (Europhysics Letters)} \textbf{\bibinfo{volume}{88}},
  \bibinfo{pages}{17012} (\bibinfo{year}{2009}).

\bibitem[{\citenamefont{Cirillo et~al.}(2011)\citenamefont{Cirillo, Mancini,
  Giuliano, and Sodano}}]{giuso3}
\bibinfo{author}{\bibfnamefont{A.}~\bibnamefont{Cirillo}},
  \bibinfo{author}{\bibfnamefont{M.}~\bibnamefont{Mancini}},
  \bibinfo{author}{\bibfnamefont{D.}~\bibnamefont{Giuliano}}, \bibnamefont{and}
  \bibinfo{author}{\bibfnamefont{P.}~\bibnamefont{Sodano}},
  \bibinfo{journal}{Nuclear Physics B} \textbf{\bibinfo{volume}{852}},
  \bibinfo{pages}{235 } (\bibinfo{year}{2011}).

\bibitem[{\citenamefont{Giuliano and Sodano}(2007)}]{giusox}
\bibinfo{author}{\bibfnamefont{D.}~\bibnamefont{Giuliano}} \bibnamefont{and}
  \bibinfo{author}{\bibfnamefont{P.}~\bibnamefont{Sodano}},
  \bibinfo{journal}{Nuclear Physics B} \textbf{\bibinfo{volume}{770}},
  \bibinfo{pages}{332 } (\bibinfo{year}{2007}).

\bibitem[{\citenamefont{Tsvelik}(2013)}]{tsve1}
\bibinfo{author}{\bibfnamefont{A.~M.} \bibnamefont{Tsvelik}},
  \bibinfo{journal}{Phys. Rev. Lett.} \textbf{\bibinfo{volume}{110}},
  \bibinfo{pages}{147202} (\bibinfo{year}{2013}).

\bibitem[{\citenamefont{Altland et~al.}(2014)\citenamefont{Altland, B\'eri,
  Egger, and Tsvelik}}]{tsve2}
\bibinfo{author}{\bibfnamefont{A.}~\bibnamefont{Altland}},
  \bibinfo{author}{\bibfnamefont{B.}~\bibnamefont{B\'eri}},
  \bibinfo{author}{\bibfnamefont{R.}~\bibnamefont{Egger}}, \bibnamefont{and}
  \bibinfo{author}{\bibfnamefont{A.~M.} \bibnamefont{Tsvelik}},
  \bibinfo{journal}{Phys. Rev. Lett.} \textbf{\bibinfo{volume}{113}},
  \bibinfo{pages}{076401} (\bibinfo{year}{2014}).

\bibitem[{\citenamefont{Tsvelik}(2014)}]{tsve3}
\bibinfo{author}{\bibfnamefont{A.~M.} \bibnamefont{Tsvelik}},
  \bibinfo{journal}{New Journal of Physics} \textbf{\bibinfo{volume}{16}},
  \bibinfo{pages}{033003} (\bibinfo{year}{2014}).

\bibitem[{\citenamefont{Giuliano
  et~al.}(2016{\natexlab{a}})\citenamefont{Giuliano, Sodano, Tagliacozzo, and
  Trombettoni}}]{gstt}
\bibinfo{author}{\bibfnamefont{D.}~\bibnamefont{Giuliano}},
  \bibinfo{author}{\bibfnamefont{P.}~\bibnamefont{Sodano}},
  \bibinfo{author}{\bibfnamefont{A.}~\bibnamefont{Tagliacozzo}},
  \bibnamefont{and}
  \bibinfo{author}{\bibfnamefont{A.}~\bibnamefont{Trombettoni}},
  \bibinfo{journal}{Nuclear Physics B} \textbf{\bibinfo{volume}{909}},
  \bibinfo{pages}{135 } (\bibinfo{year}{2016}{\natexlab{a}}).

\bibitem[{\citenamefont{Giuliano
  et~al.}(2016{\natexlab{b}})\citenamefont{Giuliano, Campagnano, and
  Tagliacozzo}}]{gct}
\bibinfo{author}{\bibfnamefont{D.}~\bibnamefont{Giuliano}},
  \bibinfo{author}{\bibfnamefont{G.}~\bibnamefont{Campagnano}},
  \bibnamefont{and}
  \bibinfo{author}{\bibfnamefont{A.}~\bibnamefont{Tagliacozzo}},
  \bibinfo{journal}{The European Physical Journal B}
  \textbf{\bibinfo{volume}{89}}, \bibinfo{pages}{251}
  (\bibinfo{year}{2016}{\natexlab{b}}).

\bibitem[{\citenamefont{{Ran} et~al.}(2017)\citenamefont{{Ran}, {Peng}, {Su},
  and {Lewenstein}}}]{2017arXiv170707838R}
\bibinfo{author}{\bibfnamefont{S.-J.} \bibnamefont{{Ran}}},
  \bibinfo{author}{\bibfnamefont{C.}~\bibnamefont{{Peng}}},
  \bibinfo{author}{\bibfnamefont{G.}~\bibnamefont{{Su}}}, \bibnamefont{and}
  \bibinfo{author}{\bibfnamefont{M.}~\bibnamefont{{Lewenstein}}},
  \bibinfo{journal}{ArXiv e-prints}  (\bibinfo{year}{2017}),
  \eprint{1707.07838}.

\bibitem[{\citenamefont{Knap et~al.}(2012)\citenamefont{Knap, Shashi, Nishida,
  Imambekov, Abanin, and Demler}}]{knap}
\bibinfo{author}{\bibfnamefont{M.}~\bibnamefont{Knap}},
  \bibinfo{author}{\bibfnamefont{A.}~\bibnamefont{Shashi}},
  \bibinfo{author}{\bibfnamefont{Y.}~\bibnamefont{Nishida}},
  \bibinfo{author}{\bibfnamefont{A.}~\bibnamefont{Imambekov}},
  \bibinfo{author}{\bibfnamefont{D.~A.} \bibnamefont{Abanin}},
  \bibnamefont{and} \bibinfo{author}{\bibfnamefont{E.}~\bibnamefont{Demler}},
  \bibinfo{journal}{Phys. Rev. X} \textbf{\bibinfo{volume}{2}},
  \bibinfo{pages}{041020} (\bibinfo{year}{2012}).

\bibitem[{\citenamefont{Schir\`{o} and Mitra}(2014)}]{schiro}
\bibinfo{author}{\bibfnamefont{M.}~\bibnamefont{Schir\`{o}}} \bibnamefont{and}
  \bibinfo{author}{\bibfnamefont{A.}~\bibnamefont{Mitra}},
  \bibinfo{journal}{Phys. Rev. Lett.} \textbf{\bibinfo{volume}{112}},
  \bibinfo{pages}{246401} (\bibinfo{year}{2014}).

\bibitem[{\citenamefont{Sindona et~al.}(2013)\citenamefont{Sindona, Goold,
  Lo~Gullo, Lorenzo, and Plastina}}]{sindona1}
\bibinfo{author}{\bibfnamefont{A.}~\bibnamefont{Sindona}},
  \bibinfo{author}{\bibfnamefont{J.}~\bibnamefont{Goold}},
  \bibinfo{author}{\bibfnamefont{N.}~\bibnamefont{Lo~Gullo}},
  \bibinfo{author}{\bibfnamefont{S.}~\bibnamefont{Lorenzo}}, \bibnamefont{and}
  \bibinfo{author}{\bibfnamefont{F.}~\bibnamefont{Plastina}},
  \bibinfo{journal}{Phys. Rev. Lett.} \textbf{\bibinfo{volume}{111}},
  \bibinfo{pages}{165303} (\bibinfo{year}{2013}).

\bibitem[{\citenamefont{Sindona et~al.}(2014)\citenamefont{Sindona, Lo~Gullo,
  Goold, and Plastina}}]{sindona2}
\bibinfo{author}{\bibfnamefont{A.}~\bibnamefont{Sindona}},
  \bibinfo{author}{\bibfnamefont{N.}~\bibnamefont{Lo~Gullo}},
  \bibinfo{author}{\bibfnamefont{J.}~\bibnamefont{Goold}}, \bibnamefont{and}
  \bibinfo{author}{\bibfnamefont{F.}~\bibnamefont{Plastina}},
  \bibinfo{journal}{New J. Phys.} \textbf{\bibinfo{volume}{16}},
  \bibinfo{pages}{045013} (\bibinfo{year}{2014}).

\bibitem[{\citenamefont{Elliott and Johnson}(2016)}]{elliott}
\bibinfo{author}{\bibfnamefont{T.~J.} \bibnamefont{Elliott}} \bibnamefont{and}
  \bibinfo{author}{\bibfnamefont{T.~H.} \bibnamefont{Johnson}},
  \bibinfo{journal}{Phys. Rev. A} \textbf{\bibinfo{volume}{93}},
  \bibinfo{pages}{043612} (\bibinfo{year}{2016}).

\bibitem[{\citenamefont{Mitchison et~al.}(2016)\citenamefont{Mitchison,
  Johnson, and Jaksch}}]{mitchison}
\bibinfo{author}{\bibfnamefont{M.~T.} \bibnamefont{Mitchison}},
  \bibinfo{author}{\bibfnamefont{T.~H.} \bibnamefont{Johnson}},
  \bibnamefont{and} \bibinfo{author}{\bibfnamefont{D.}~\bibnamefont{Jaksch}},
  \bibinfo{journal}{Phys. Rev. A} \textbf{\bibinfo{volume}{94}},
  \bibinfo{pages}{063618} (\bibinfo{year}{2016}).

\bibitem[{\citenamefont{Tamascelli et~al.}(2016)\citenamefont{Tamascelli,
  Benedetti, Olivares, and Paris}}]{tamascelli}
\bibinfo{author}{\bibfnamefont{D.}~\bibnamefont{Tamascelli}},
  \bibinfo{author}{\bibfnamefont{C.}~\bibnamefont{Benedetti}},
  \bibinfo{author}{\bibfnamefont{S.}~\bibnamefont{Olivares}}, \bibnamefont{and}
  \bibinfo{author}{\bibfnamefont{M.~G.~A.} \bibnamefont{Paris}},
  \bibinfo{journal}{Phys. Rev. A} \textbf{\bibinfo{volume}{94}},
  \bibinfo{pages}{042129} (\bibinfo{year}{2016}).

\bibitem[{\citenamefont{Streif et~al.}(2016)\citenamefont{Streif, Buchleitner,
  Jaksch, and Mur-Petit}}]{streif}
\bibinfo{author}{\bibfnamefont{M.}~\bibnamefont{Streif}},
  \bibinfo{author}{\bibfnamefont{A.}~\bibnamefont{Buchleitner}},
  \bibinfo{author}{\bibfnamefont{D.}~\bibnamefont{Jaksch}}, \bibnamefont{and}
  \bibinfo{author}{\bibfnamefont{J.}~\bibnamefont{Mur-Petit}},
  \bibinfo{journal}{Phys. Rev. A} \textbf{\bibinfo{volume}{94}},
  \bibinfo{pages}{053634} (\bibinfo{year}{2016}).

\bibitem[{\citenamefont{Cosco et~al.}(2017)\citenamefont{Cosco, Borrelli,
  Plastina, and Maniscalco}}]{plastina_1}
\bibinfo{author}{\bibfnamefont{F.}~\bibnamefont{Cosco}},
  \bibinfo{author}{\bibfnamefont{M.}~\bibnamefont{Borrelli}},
  \bibinfo{author}{\bibfnamefont{F.}~\bibnamefont{Plastina}}, \bibnamefont{and}
  \bibinfo{author}{\bibfnamefont{S.}~\bibnamefont{Maniscalco}},
  \bibinfo{journal}{Phys. Rev. A} \textbf{\bibinfo{volume}{95}},
  \bibinfo{pages}{053620} (\bibinfo{year}{2017}).

\bibitem[{\citenamefont{Peschel}(1984)}]{peschel_1}
\bibinfo{author}{\bibfnamefont{I.}~\bibnamefont{Peschel}},
  \bibinfo{journal}{Phys. Rev. B} \textbf{\bibinfo{volume}{30}},
  \bibinfo{pages}{6783} (\bibinfo{year}{1984}).

\bibitem[{\citenamefont{Fradkin and Susskind}(1978)}]{fradkin_78}
\bibinfo{author}{\bibfnamefont{E.}~\bibnamefont{Fradkin}} \bibnamefont{and}
  \bibinfo{author}{\bibfnamefont{L.}~\bibnamefont{Susskind}},
  \bibinfo{journal}{Phys. Rev. D} \textbf{\bibinfo{volume}{17}},
  \bibinfo{pages}{2637} (\bibinfo{year}{1978}).

\bibitem[{\citenamefont{Kogut}(1979)}]{kogut_79}
\bibinfo{author}{\bibfnamefont{J.~B.} \bibnamefont{Kogut}},
  \bibinfo{journal}{Rev. Mod. Phys.} \textbf{\bibinfo{volume}{51}},
  \bibinfo{pages}{659} (\bibinfo{year}{1979}).

\bibitem[{\citenamefont{Um et~al.}(2007)\citenamefont{Um, Lee, and Kim}}]{kim}
\bibinfo{author}{\bibfnamefont{J.}~\bibnamefont{Um}},
  \bibinfo{author}{\bibfnamefont{S.-I.} \bibnamefont{Lee}}, \bibnamefont{and}
  \bibinfo{author}{\bibfnamefont{B.~J.} \bibnamefont{Kim}},
  \bibinfo{journal}{Journal of the Korean Physical Society}
  \textbf{\bibinfo{volume}{50}}, \bibinfo{pages}{285} (\bibinfo{year}{2007}).

\bibitem[{\citenamefont{Burkhardt and Guim}(1985)}]{0305-4470-18-1-006}
\bibinfo{author}{\bibfnamefont{T.~W.} \bibnamefont{Burkhardt}}
  \bibnamefont{and} \bibinfo{author}{\bibfnamefont{I.}~\bibnamefont{Guim}},
  \bibinfo{journal}{Journal of Physics A: Mathematical and General}
  \textbf{\bibinfo{volume}{18}}, \bibinfo{pages}{L33} (\bibinfo{year}{1985}).

\bibitem[{\citenamefont{Lieb et~al.}(1961)\citenamefont{Lieb, Schultz, and
  Mattis}}]{LIEB1961407}
\bibinfo{author}{\bibfnamefont{E.}~\bibnamefont{Lieb}},
  \bibinfo{author}{\bibfnamefont{T.}~\bibnamefont{Schultz}}, \bibnamefont{and}
  \bibinfo{author}{\bibfnamefont{D.}~\bibnamefont{Mattis}},
  \bibinfo{journal}{Annals of Physics} \textbf{\bibinfo{volume}{16}},
  \bibinfo{pages}{407 } (\bibinfo{year}{1961}).

\bibitem[{\citenamefont{Barouch and McCoy}(1971)}]{Barouch2}
\bibinfo{author}{\bibfnamefont{E.}~\bibnamefont{Barouch}} \bibnamefont{and}
  \bibinfo{author}{\bibfnamefont{B.~M.} \bibnamefont{McCoy}},
  \bibinfo{journal}{Phys. Rev. A} \textbf{\bibinfo{volume}{3}},
  \bibinfo{pages}{786} (\bibinfo{year}{1971}).

\bibitem[{\citenamefont{Onsager}(1944)}]{PhysRev.65.117}
\bibinfo{author}{\bibfnamefont{L.}~\bibnamefont{Onsager}},
  \bibinfo{journal}{Phys. Rev.} \textbf{\bibinfo{volume}{65}},
  \bibinfo{pages}{117} (\bibinfo{year}{1944}).

\bibitem[{\citenamefont{Yang}(1952)}]{PhysRev.85.808}
\bibinfo{author}{\bibfnamefont{C.~N.} \bibnamefont{Yang}},
  \bibinfo{journal}{Phys. Rev.} \textbf{\bibinfo{volume}{85}},
  \bibinfo{pages}{808} (\bibinfo{year}{1952}).

\bibitem[{\citenamefont{Dutta et~al.}(2015)\citenamefont{Dutta, Aeppli,
  Chakrabarti, Divakaran, Rosenbaum, and Sen}}]{book:1390478}
\bibinfo{author}{\bibfnamefont{A.}~\bibnamefont{Dutta}},
  \bibinfo{author}{\bibfnamefont{G.}~\bibnamefont{Aeppli}},
  \bibinfo{author}{\bibfnamefont{B.~K.} \bibnamefont{Chakrabarti}},
  \bibinfo{author}{\bibfnamefont{U.}~\bibnamefont{Divakaran}},
  \bibinfo{author}{\bibfnamefont{T.~F.} \bibnamefont{Rosenbaum}},
  \bibnamefont{and} \bibinfo{author}{\bibfnamefont{D.}~\bibnamefont{Sen}},
  \emph{\bibinfo{title}{Quantum Phase Transitions in Transverse Field Spin
  Models: From Statistical Physics to Quantum Information}}
  (\bibinfo{publisher}{Cambridge University Press}, \bibinfo{year}{2015}).

\bibitem[{\citenamefont{Griffiths}(1969)}]{PhysRevLett.23.17}
\bibinfo{author}{\bibfnamefont{R.~B.} \bibnamefont{Griffiths}},
  \bibinfo{journal}{Phys. Rev. Lett.} \textbf{\bibinfo{volume}{23}},
  \bibinfo{pages}{17} (\bibinfo{year}{1969}).

\bibitem[{\citenamefont{Vojta}(2003)}]{PhysRevLett.90.107202}
\bibinfo{author}{\bibfnamefont{T.}~\bibnamefont{Vojta}},
  \bibinfo{journal}{Phys. Rev. Lett.} \textbf{\bibinfo{volume}{90}},
  \bibinfo{pages}{107202} (\bibinfo{year}{2003}).

\bibitem[{\citenamefont{Vojta}(2014)}]{1742-6596-529-1-012016}
\bibinfo{author}{\bibfnamefont{T.}~\bibnamefont{Vojta}},
  \bibinfo{journal}{Journal of Physics: Conference Series}
  \textbf{\bibinfo{volume}{529}}, \bibinfo{pages}{012016}
  (\bibinfo{year}{2014}).

\bibitem[{\citenamefont{M\"uller and Nersesyan}(2016)}]{nersesyan}
\bibinfo{author}{\bibfnamefont{M.}~\bibnamefont{M\"uller}} \bibnamefont{and}
  \bibinfo{author}{\bibfnamefont{A.~A.} \bibnamefont{Nersesyan}},
  \bibinfo{journal}{Annals of Physics} \textbf{\bibinfo{volume}{372}},
  \bibinfo{pages}{482} (\bibinfo{year}{2016}).

\bibitem[{\citenamefont{Francica et~al.}(2016)\citenamefont{Francica, Apollaro,
  Lo~Gullo, and Plastina}}]{apollaro}
\bibinfo{author}{\bibfnamefont{G.}~\bibnamefont{Francica}},
  \bibinfo{author}{\bibfnamefont{T.~J.~G.} \bibnamefont{Apollaro}},
  \bibinfo{author}{\bibfnamefont{N.}~\bibnamefont{Lo~Gullo}}, \bibnamefont{and}
  \bibinfo{author}{\bibfnamefont{F.}~\bibnamefont{Plastina}},
  \bibinfo{journal}{Phys. Rev. B} \textbf{\bibinfo{volume}{94}},
  \bibinfo{pages}{245103} (\bibinfo{year}{2016}).

\bibitem[{\citenamefont{Jordan and Wigner}(1928)}]{jordanwigner}
\bibinfo{author}{\bibfnamefont{P.}~\bibnamefont{Jordan}} \bibnamefont{and}
  \bibinfo{author}{\bibfnamefont{E.}~\bibnamefont{Wigner}},
  \bibinfo{journal}{Z. Phys.} \textbf{\bibinfo{volume}{47}},
  \bibinfo{pages}{531} (\bibinfo{year}{1928}).

\bibitem[{\citenamefont{Schultz et~al.}(1964)\citenamefont{Schultz, Mattis, and
  Lieb}}]{RevModPhys.36.856}
\bibinfo{author}{\bibfnamefont{T.~D.} \bibnamefont{Schultz}},
  \bibinfo{author}{\bibfnamefont{D.~C.} \bibnamefont{Mattis}},
  \bibnamefont{and} \bibinfo{author}{\bibfnamefont{E.~H.} \bibnamefont{Lieb}},
  \bibinfo{journal}{Rev. Mod. Phys.} \textbf{\bibinfo{volume}{36}},
  \bibinfo{pages}{856} (\bibinfo{year}{1964}).

\bibitem[{\citenamefont{McCoy and Wu}(1973)}]{mccoy}
\bibinfo{author}{\bibfnamefont{B.}~\bibnamefont{McCoy}} \bibnamefont{and}
  \bibinfo{author}{\bibfnamefont{T.}~\bibnamefont{Wu}},
  \emph{\bibinfo{title}{The two-dimensional Ising Model}}
  (\bibinfo{publisher}{Harvard University Press, Cambridge},
  \bibinfo{year}{1973}).

\bibitem[{\citenamefont{Pfeuty}(1970)}]{PFEUTY197079}
\bibinfo{author}{\bibfnamefont{P.}~\bibnamefont{Pfeuty}},
  \bibinfo{journal}{Annals of Physics} \textbf{\bibinfo{volume}{57}},
  \bibinfo{pages}{79 } (\bibinfo{year}{1970}).

\bibitem[{\citenamefont{Domb et~al.}(1983)\citenamefont{Domb, Green, and
  Lebowitz}}]{domb1983phase}
\bibinfo{author}{\bibfnamefont{C.}~\bibnamefont{Domb}},
  \bibinfo{author}{\bibfnamefont{M.}~\bibnamefont{Green}}, \bibnamefont{and}
  \bibinfo{author}{\bibfnamefont{J.}~\bibnamefont{Lebowitz}},
  \emph{\bibinfo{title}{Phase transitions and critical phenomena}}, no.
  \bibinfo{number}{v. 8} in \bibinfo{series}{Phase Transitions and Critical
  Phenomena} (\bibinfo{publisher}{Academic Press}, \bibinfo{year}{1983}).

\bibitem[{\citenamefont{Igl\'{o}i and Lin}(2008)}]{1742-5468-2008-06-P06004}
\bibinfo{author}{\bibfnamefont{F.}~\bibnamefont{Igl\'{o}i}} \bibnamefont{and}
  \bibinfo{author}{\bibfnamefont{Y.-C.} \bibnamefont{Lin}},
  \bibinfo{journal}{Journal of Statistical Mechanics: Theory and Experiment}
  \textbf{\bibinfo{volume}{2008}}, \bibinfo{pages}{P06004}
  (\bibinfo{year}{2008}).

\bibitem[{\citenamefont{Igl\'oi et~al.}(1993)\citenamefont{Igl\'oi, Peschel,
  and Turban}}]{peschel_2}
\bibinfo{author}{\bibfnamefont{F.}~\bibnamefont{Igl\'oi}},
  \bibinfo{author}{\bibfnamefont{I.}~\bibnamefont{Peschel}}, \bibnamefont{and}
  \bibinfo{author}{\bibfnamefont{L.}~\bibnamefont{Turban}},
  \bibinfo{journal}{Advances in Physics} \textbf{\bibinfo{volume}{42}},
  \bibinfo{pages}{683} (\bibinfo{year}{1993}).

\end{thebibliography}

%


\end{document}